\documentclass[12pt, draftclsnofoot, onecolumn]{IEEEtran}
\usepackage{amsthm,amsmath,amssymb,mathtools,bm,etoolbox,float}
\usepackage[dvipsnames]{xcolor}
\usepackage{graphicx,cite,cuted}
\usepackage{bigints}
\usepackage{caption}
\newcommand{\e}[1]{{\mathbb E}\left[ #1 \right]}
\newcommand\ddfrac[2]{\frac{\displaystyle #1}{\displaystyle #2}}

\usepackage{stackengine}

\usepackage{mathrsfs,verbatim}
\usepackage{ellipsis}
\usepackage{tkz-euclide}
\usepackage{tikz}
\usepackage{tikz-3dplot}
\usepackage[utf8]{inputenc}
\usepackage{pgfplots} 
\usepackage{pgfgantt}
\usepackage{pdflscape}
\pgfplotsset{compat=newest} 
\pgfplotsset{plot coordinates/math parser=false}
\pgfplotsset{every  tick/.style={black,},ylabel style={font=\tiny},xlabel style={font=\tiny},tick label style={font=\tiny},legend style= {font=\scriptsize},
minor x tick num=1,minor y tick num=1,xminorticks=true,yminorticks=true,}
  \newlength\fheight
\newlength\fwidth

\usepackage{xinttools} 
\usepackage{tikz}
\usetikzlibrary{calc}
\newtheorem{remark}{Remark}
\newtheorem{theorem}{Theorem}
\newtheorem{corollary}{Corollary}

\def\biglen{20cm} 
\tikzset{
  half plane/.style={ to path={
       ($(\tikztostart)!.5!(\tikztotarget)!#1!(\tikztotarget)!\biglen!90:(\tikztotarget)$)
    -- ($(\tikztostart)!.5!(\tikztotarget)!#1!(\tikztotarget)!\biglen!-90:(\tikztotarget)$)
    -- ([turn]0,2*\biglen) -- ([turn]0,2*\biglen) -- cycle}},
  half plane/.default={1pt}
}

\DeclareMathAlphabet{\pazocal}{OMS}{zplm}{m}{n}

\usepackage{caption}
\usepackage{multicol,subcaption}
\begin{document}

\title{Stochastic Geometry Analysis of Uplink Cellular Networks with FSO Backhauling: Cooperative Relaying Vs. Reflecting Surfaces}

\author{Elyes~Balti,~\IEEEmembership{Student~Member,~IEEE,} and
        Brian~K.~Johnson,~\IEEEmembership{Senior~Member,~IEEE}}

\maketitle

\begin{abstract}
In this work, we consider the performance analysis of the uplink cellular networks with free space optics (FSO) backhauling. The user equipment (UE) communicates with the nearest Base Station (BS) in first slot while in second slot, the BS converts the received radio frequency (RF) signal into FSO pulse and transmits to the data center. We adopt the Rayleigh fading for the uplink channels while the FSO backhaul encompasses the turbulence-induced fading which follows the M\'alaga distribution, the weather pathloss, and the pointing errors fading which is distributed following the generalized Beckmann model. Further, we will compare the performances when the BS behaves as either a decode-and-forward (DF) relay or an intelligent reflecting surface (IRS). Next, we will propose an optimal design of the phase shifters of the IRS to minimize the interference and improve the rate to beat relaying. Capitalizing on this framework, we will derive the system performance metrics such as the coverage probability as well as the spectral efficiency. Focusing on high SNR, we will obtain the diversity gain to get engineering insights into the system performance and limitations. Finally, the analytical results are confirmed by Monte Carlo simulations.
\end{abstract}

\IEEEpeerreviewmaketitle

\section{Introduction}
\IEEEPARstart{M}{odern} radio cellular systems are evolving from second generation (2G), where the networks are voice-oriented, to 4G-LTE (Long-Term Evolution), where the networks support medias, social networks and videos. For LTE-A (Advanced), the maximum data rate is limited to 100 Mbits/s for high mobility while typically much lower rates are achieved \cite{v1,v2}. 

Due to the increasing growth of the subscribers and data size, conventional backhauling such as microwave radio-frequency (RF) which represents 6\% of total usage after the copper lines in US, become unable to meet such requirements \cite{cu}. Recently, related works proposed as an alternative the optical fibers (OF) to backhaul the information from the base stations (BSs) to the data center \cite{surv}. Such technique has gained enormous attention since it can support the exchange of big data and reduces the congestion. In addition, OF provides not only high data rates and immunity to the interference, but also low probability of error and supports reliable communications over long distances, e.g., 155.52 Mbit/s for STM-1, 622 Mbit/s for STM-4, 2.4 Gbit/s for STM-16 and 9.9 Gbit/s for STM-64 \cite{cu}. However, OF cannot be deployed in any area and the total usage is still limited since it is very expensive. Moreover, OF are not flexible as in some cases, when the transmitter (TX) and receiver (RX) are not in line-of-sight, bending is required to establish the communication \cite{trinh}. However, studies showed that bending introduced severe losses to the signal propagating in OF \cite{bend}. The features make the total usage of OF in US below 4\% \cite{of4}.

To overtake these limitations, free-space optical (FSO) communication has been proposed as an alternative or complementary to RF and OF to cover the networks backhauling \cite{uysal,nasab,zedini}. Unlike the aforementioned backhauling solutions, FSO technology is cost-efficient, highly secure, free-license to spectrum access and flexible to be deployed in any areas without restrictions. These properties make the FSO to exponentially increases the networks coverage and data rate 25 fold more efficient than conventional backhauling and it is mainly used to densify the networks and meet the requirements of the increasing subscribers over the last decades \cite{d1,d2,d3}. Furthermore, FSO has been introduced in applications related to academia and industries such as videos surveillance, disaster recovery, broadcasting, enterprise/campus connectivities and redundant links \cite{a1,surv,uysal}. 

\subsection{FSO Technology Overview}
As mentioned earlier, FSO has been emerged as promising solutions for many wireless systems issues, in particular for backhauling where it becomes a reliable solution for the last mile problem to bridge the bandwidth gap between the end-users and the OF backbone network \cite{back1,back2,back3,antenna}. However, FSO technology has also some limitations that were the main focus of recent works. In fact, FSO links are established at high altitudes on the top of buildings to protect people from laser vulnerabilities and avoid potential eyes damaging \cite{saf1,saf2}. At a high altitude, the channels are extremely turbulent and feature deep fading resulting in severe losses to FSO signaling. Previous works extensively examined probabilistic channels models to describe the atmospheric turbulences variations. Lognormal, Gamma-Gamma and Double-Weibull distributions have been proposed to model the turbulences-induced fading \cite{farid,icc,glob16,glob17}. These models were shown to be valid only for weak and moderate turbulences severity, however, they exhibit large deviation from experimental data for strong turbulences \cite{limit1,limit2}. Recent works addressed this shortcoming by suggesting generalized models such that the double generalized gamma and M\'alaga that both reflect wide range of turbulences including the severe fading \cite{limit2,mal1,mal2}. 
During the propagation, FSO signaling is subject to the pointing errors caused by the misalignment between the laser-emitting transmitter and the photodetector. External factors like buildings swaying and seismic activities are potentially originating the misalignment \cite{surv,uysal}. In fact, pointing errors introduce random fluctuations to the FSO signaling to the point that the photodetector cannot intercept the total received power resulting in performance degradation. To quantify these losses, existing work proposed different fading models to characterize the pointing errors severity and impacts on the system performance \cite{surv,uysal}. The most generalized pointing errors is called Beckmann model \cite{beck} where it reflects various special cases to model the radial displacement such as Hoyt \cite{p2}, Rician \cite{p1}, Zero-Mean and Non Zero-Mean Single-Sided Gaussian \cite{p3}, while the widely used model is Rayleigh for tractability \cite{tract,mixed,aggregate}. 

\subsection{Relaying and Intelligent Reflecting Surfaces Overview}       
Due to the big cell size, macrocellular systems exhibit low coverage specifically the cell-edge users who are exposed to the severe pathloss and near-far effect. Relaying has been proposed as a solution to increase the network scalability and improve the coverage \cite{ibo,tract,mixed,aggregate,v2xdiversity,asym,eT}. In this context, related research attempts developed various relaying schemes, e.g., Decode-and-Forward (DF) \cite{tract,mixed,joint}, Amplify-and-Forward (AF) for fixed and variable relaying gains \cite{glob16,variable,ibo}, Quantize-and-Encode (QE) \cite{qe} and Quantize-and-Forward (QF) \cite{qf}. Another attractive solution, called intelligent reflecting surfaces (IRS), has been proposed to address the coverage, scalability and energy consumption issues \cite{irs1,irs2,irs3}. Such IRS consists of a single or few-layer stack of planar structures that are fabricated using lithography and nano-printing methods.
 IRSs are simply meta-surfaces equipped with integrated electronic circuits that can be programmed to reflect the incoming electromagnetic waves in a specific direction. Each reflecting element employs varactor diodes or other micro electro-mechanical systems whose resonant frequency is manageable to control the direction of the reflected signal. Unlike the parabolic reflectors whose physical curvature determine the desired direction of the reflected signal, IRS is flat and consists of an array of discrete reflecting elements that phase-shift and beamform the incident waves differently. The phase shifter of each element determines in which direction the incident waves are beamformed and hence an efficient configuration and programming of the reflecting elements is required. Unlike relaying, IRS is a passive reflector component and it does not containt a power amplifier such that for amplify-and-forward relaying. Thereby, from an energy consumption perspective, IRS has been shown to be more energy efficient and it consumes less power compared to traditional relaying. From a cost-efficiency standpoint, however, IRS is very expensive compared to relaying and as a result such a magic solution cannot be deployed anywhere. For outdoor communications wherein IRSs have to deployed in companies ceilings and building facades across urban street canyon and highways, a well-established trade-off between performance and cost favoritize relaying over IRS. Though, IRS is still practical, in particular, for indoor and more importantly in tunnels wherein GPS service is absent and consequently IRSs can be optimally deployed in a smart way to maintain affordable costs and improve the coverage in such places. Results provided by \cite{14,26} demonstrated that IRS can achieve amplification gains to support the received signal while it substantially improves the signal-to-interference-plus-noise-ratio (SINR) by designing the phase shifters to cancel the interference. In addition, indoor measurements in office testbed were conducted to corroborate the results for RIS deployment while , \cite{14,26} carried out rigorous analysis on IRS roles to enhance the indoor coverage. Furthermore, \cite{27} experiments were driven on the deployment of IRS in an IEEE 802.11ad network operating in the unlicensed 60 GHz spectrum. Moreover, application of IRS for actively reprogramming communication environments was tackled and its impacts on the energy consumption, coverage and security concerns were analyzed. Besides, research attempts have introduced the full-duplex relaying as it has the potential to double the spectral efficiency. Due to the self-interference which substantially degrades the performance, related works proposed beamforming designs to cancel the self-interference and improve the achievable rate \cite{fd1,fd2}. Furthermore, relaying as well as the intelligent reflecting surfaces (IRS) have been introduced in the context of physical layer security of vehicular adhoc networks (VANET) to protect the legitimate receiver from the eavesdropping attacks by transmitting friendly jammers and/or artificial noise in order to maximize the secrecy capacity \cite{pls1,pls2,neji1,neji2,maalej}.
 \subsection{Contributions}
 In this work, we consider the uplink of an outdoor cellular radio network in which a user equipment (UE) reaches to a data center through a base station (BS). FSO laser signaling takes place between the BS and the data center in line-of-sight (LOS) at high altitude to protect the people around from potential dangers caused by laser beam. We assume Rayleigh fading to model the uplink channels where the fractional power control (FPC) is considered to manage the pathloss severity on the UE. Further, the FSO backhauling encompasses the pathloss, the pointing errors distributed following the generalized Beckmann model and the turbulence-induced fading that follows the M\'alaga or $\mathcal{M}$ probabilistic model. Moreover, we provide a global framework analysis of two generic scenarios in which the BS behaves as a relay or IRS between the uplink UE and the data center. Given that DF is known to outperform AF relaying in terms of coverage and rate, we consider DF as a benchmark to provide a fair comparison with IRS-supported transmission. The objective is to provide a phase shifters design for IRS to determine the minimum number of reflecting element required to beat repetition-coded DF relaying. To this end, the paper analysis follows these steps:
 \begin{itemize}
 \item Using stochastic geometry tools, we derive the tractable expressions of the coverage/outage and ergodic achievable rate based on the statistics of the signal-to-interference-plus-noise-ratio (SINR) of a typical uplink UE.
 \item Derive the statistics of the FSO backhauling channels involving the probability density function (PDF), cumulative distribution function (CDF), high order moments, exact, low/high asymptotic SNR and the upper bound of the achievable rate.
 \item Exploiting the results obtained above, we derive the  coverage/outage and rate expressions for the repetition-coded DF relaying.
 \item To design the optimal phase shifters of the IRS, we formulate the optimization problem and construct the zero-forcing constraint to maximize the end-to-end SINR which is the objective function of such a problem.
 \end{itemize}
 \subsection{Structure}
 This paper is organized as follows: Section II describes the uplink cellular system along with the derivation of the coverage probability and the ergodic rate for an uplink typical user using stochastic geometry tool. The performance of FSO backhauling analysis is detailed in Section III while the analysis of the hybrid system involving the analysis of repetition-coded DF relaying and IRS-supported transmission is presented in Section IV. Section V presents the numerical results along with the discussion while the concluding remarks are reported in VI.
  \subsection{Notation}
  For the sake of organization, we provide some useful notations to avoid repetition. $F_h(\cdot)$ and $f_h(\cdot)$ are the CDF and PDF of the random variables $h$, respectively. $\mathbb{E}[\cdot]$ and $\mathbb{P}[\cdot]$ denote the expectation operator and the probability measure, respectively.
\section{Cellular Networks Analysis}
Recent researches on modern cellular networks proposed various approaches to address the uplink modeling and performance analysis. Wyner model, proposed in \cite{2}, gained enormous attention for its simplicity as it consider inter-cell interference to be constant or a single random variable to account for fading. However, it has been shown in \cite{9} that this model suffers from shortcoming in the cases where the inter-cell interference is spatially averaged, e.g., a code division multiple access (CDMA) networks under high load congestion. In the same context, existing work suggested a network model consisting of a regular and deterministic BSs deployment, e.g., hexagonal or grid-based model shown by Fig.~\ref{a1}. Although this model is regular, closed-form expressions of the performance are not tractable and instead the derivation is based on approximation and Monte Carlo simulation. It has been shown that the grid-based model is an upper bound to the actual performance retrieved from real data, e.g., the SINR coverage based upon hexagonal model is an upper bound to a given actual network results \cite{jeff1,jeff2}. Limitations of previous approaches triggered the focus on the use of random spatial models for networks modeling. This new perspective brought two important results wherein the first consists of tractable derivation of performance such as the rate and SINR expressions while the second advantage is as accurate as grid-based model or even more closer in some cases to the actual networks, e.g., illustrated by Fig.~\ref{b1} is a Poisson Point Process (PPP) distribution of both BSs deployment and UEs provides accurate SINR statistics for real urban networks in \cite{22}. Fig.~\ref{fig2} illustrates an uplink cellular network where the UE in the cell of interest (red) is communicating with the corresponding BS which is also subject to the UEs interference coming from the adjacent cells (yellow and green). Such network consists only of the first tier (macro BSs) and the UEs are randomly distributed following a homogeneous Poisson Point Process (PPP) with density $\lambda$. 
\begin{figure}[H]
\begin{subfigure}[b]{0.5\textwidth}
\centering
\includegraphics{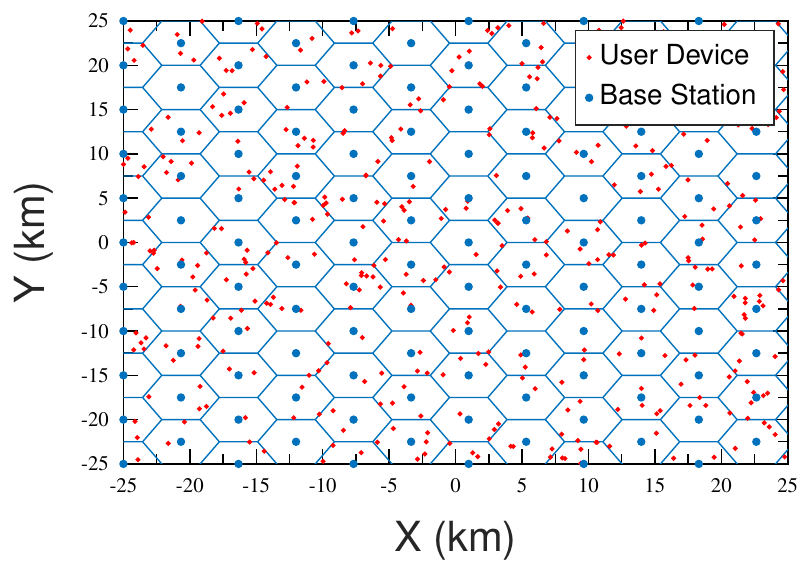}
    \caption{Hexagonal lattice.}
    \label{a1}
    \end{subfigure}
    \begin{subfigure}[b]{0.5\textwidth}
\centering
\includegraphics{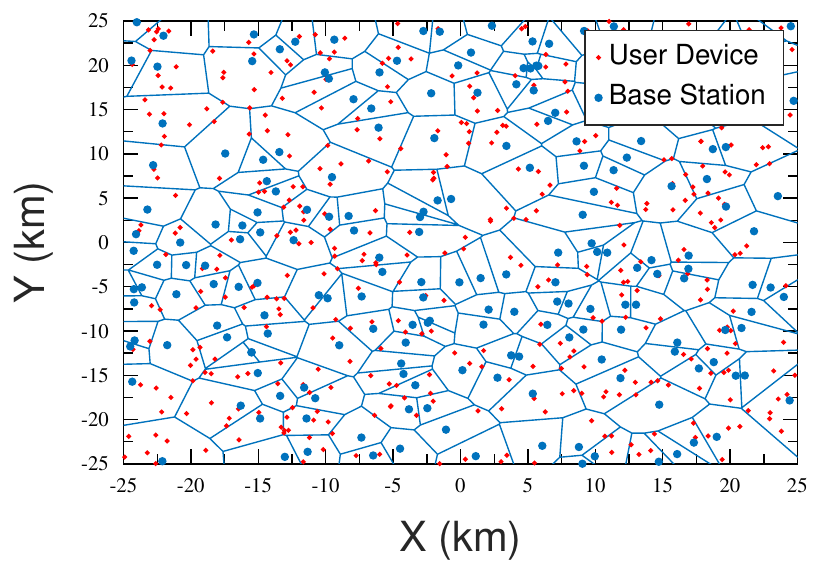}
    \caption{Voronoi tessellation.}
    \label{b1}
    \end{subfigure}
    \caption[map]{Realization of single tier cellular networks over $50 \times 50$ km$^2$ area. Hexagonal grid consists of regular deployment of BSs while the users devices are randomly distributed in the cells (a). PPP network model wherein the BSs deployment and users devices are distributed according to PPP. }
    \label{fig1}
\end{figure}
\begin{figure}[H]
    \centering
    \includegraphics[width=10cm]{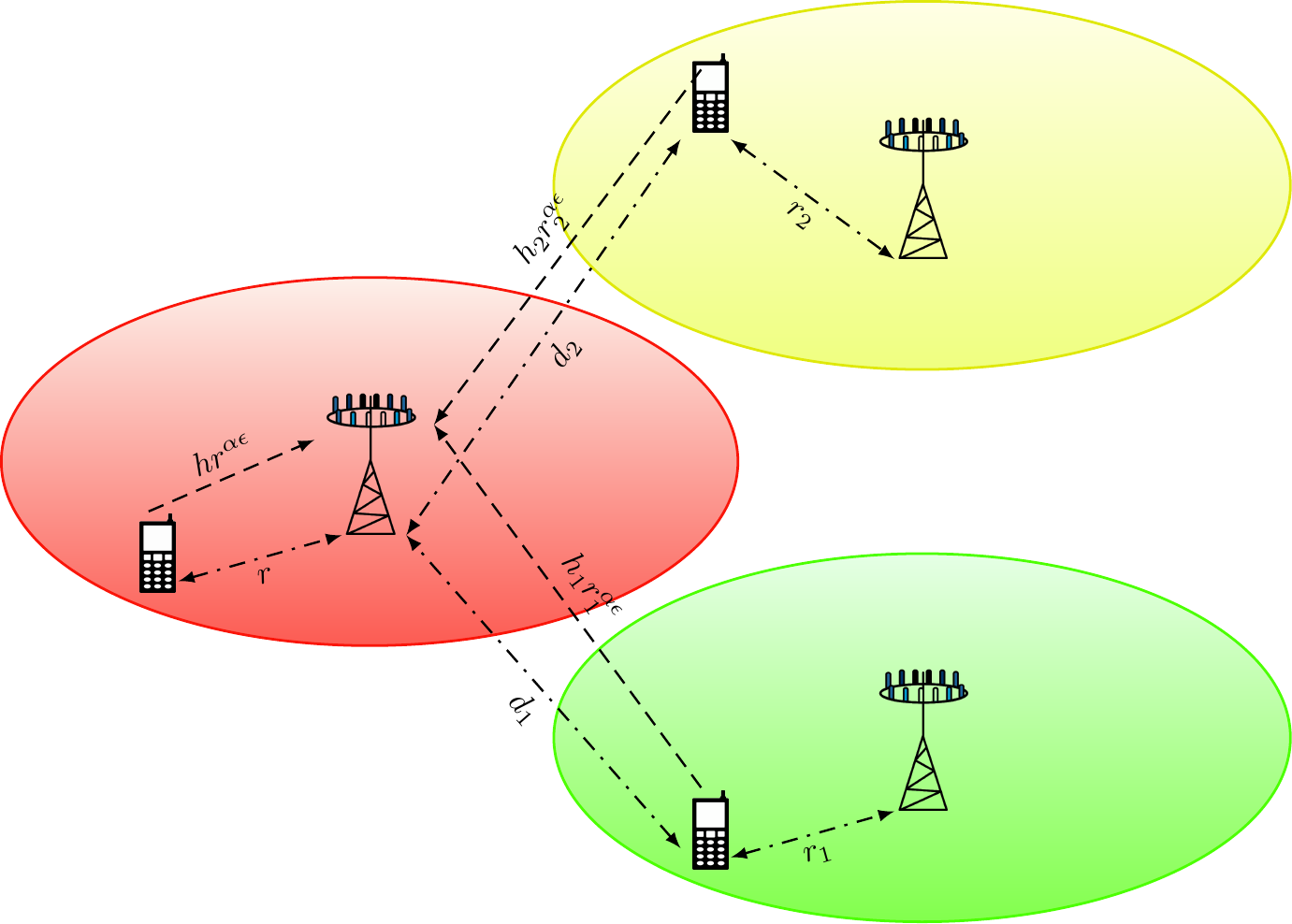}
    \caption{Scenario of uplink cellular network. The BS of interest located in the red cell is subject to the UEs interference from the other cells. The power is controlled by the fractional power control $\epsilon \in [0~1]$.}
    \label{fig2}
\end{figure}

Given that in such outdoor environment, the received power is highly degraded by the pathloss and the shadowing, we adopt a slow power regime such as the fractional power control to compensate for the pathloss. Such power adaptation technique is characterized by a parameter $\epsilon$ to control the received power of the UEs. In this context, we assume that all the UEs use the distance-proportional fractional power control as $r_z^{\alpha\epsilon}$. If $\epsilon = 0$, means that all the UEs have the same transmit power and no power control regime is adopted. When $\epsilon = 1$ means that the pathloss exponent is completely compensated.
The channels between the UEs and the BS of interest are subject to Rayleigh fading with mean power $\mu$ while the pathloss is inversely proportional to the distance with an exponent $\alpha$.

The signal-to-interference-plus-noise-ratio (SINR) at the BS of interest is given by
\begin{equation}{\label{eq1}}
{\scriptsize{\textsf{SINR}}} = \frac{hr^{\alpha(\epsilon-1)}}{\sigma^2 + \mathcal{I}_z}, 
\end{equation}

where $\sigma^2$ is the noise power and $\mathcal{I}_z$ is the total amount of interference given by
\begin{equation}{\label{eq2}}
\mathcal{I}_z = \sum_{z\in \mathcal{Z}}r_z^{\alpha\epsilon}h_zd_z^{-\alpha}.   
\end{equation}
Note that $d_z$ is the distance between the interfering UE to the BS of interest, $h_z$ is the interfering fading and $r_z$ is the distance between the interfering UE to its serving BS.
\begin{figure}[H]
\begin{subfigure}[b]{0.5\textwidth}
\centering
\includegraphics{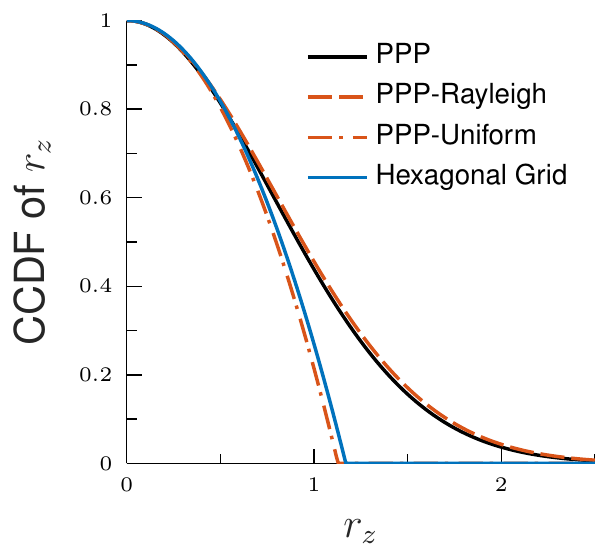}
    \caption{}
    \label{a3}
    \end{subfigure}
    \begin{subfigure}[b]{0.5\textwidth}
\centering
\includegraphics{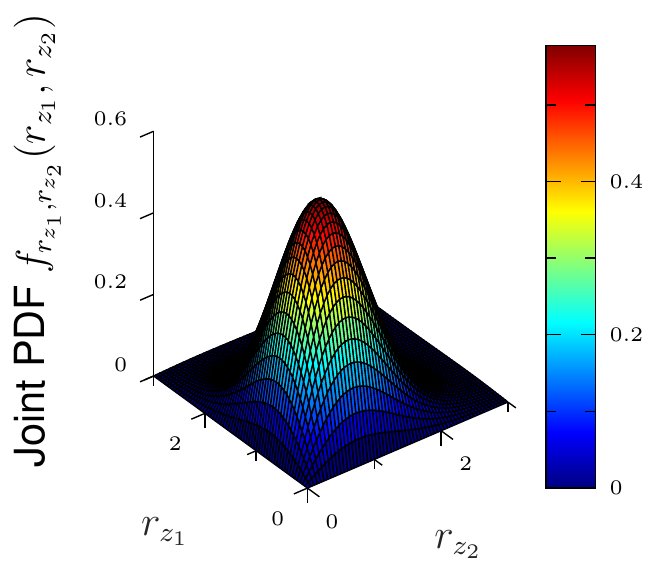}
    \caption{}
    \label{b3}
    \end{subfigure}
    \caption[map]{A comparison of the complementary CDFs (CCDFs) of $r_z$ both for the PPP and a hexagonal grid model with their respective approximation for $\lambda = \frac{1}{4}$ in (a). Joint densities of $r_{z_1}$ and $r_{z_2}$ for the actual PPP model under the independence assumption in (b). $r_{z_1}$ and $r_{z_2}$ are the distances of the UEs to their respective BSs in two adjacent Voronoi cells. The joint PDF density is $f_{r_{z_1},r_{z_2}}(r_{z_1},r_{z_2}) = (2\pi\lambda)^2r_{z_1}r_{z_2}e^{-\lambda\pi(r_{z_1}^2+r_{z_2}^2)}, r_{z_1} \geq 0,~r_{z_2}\geq 0$.   }
    \label{fig3}
\end{figure}
Various special cases on the distribution of the random variable $r_z$ can be considered to get insights into different intuitions.
\begin{itemize}
\item \textbf{PPP:} This model is general and does not consider any assumptions wherein the UEs are distributed following the spatial PPP within a Voronoi cell containing only one BS.
\item \textbf{PPP-Rayleigh:} This model follows the same setting as PPP above but it differs only on $r_z$ which is independent and identically distributed (i.i.d) as Rayleigh. The PDF of $r_z$ is $f_{r_z}(r_z) = 2\pi\lambda r_z e^{-\lambda\pi r_z^2},~ r_z \geq 0$.
\item \textbf{PPP-Uniform:} This setup differs from Rayleigh case in the distribution of $r_z$ wherein the BS is uniformly deployed in a circle around the UE. This model leads to the results relative to the uniform deployment of BSs. Each hexagonal cell is approximated as a circle having the same area $\lambda^{-1}$ and a radius of $\frac{1}{\sqrt{\pi\lambda}}$ around its corresponding UE. The PDF of $r_z$ is $f_{r_z}(r_z) = 2\pi\lambda r_z,~0 \leq r_z \leq \frac{1}{\sqrt{\pi\lambda}}$.
\item \textbf{Hexagonal:} This model consists of regular deployment of BS in the center of each hexagon and one UE is uniformly distributed in each cell. This model does not yield tractable formulations and it is simulated using exhaustive Monte Carlo iterations.
\end{itemize}

\subsection{Coverage Analysis}
For a given SINR threshold $\Gamma$, the coverage probability is defined as the probability when the SINR is greater than the target threshold. It can be written as
\begin{equation}{\label{eq5}}
P_c({\scriptsize{\textsf{SINR}}},\Gamma) = \mathbb{P}[{\scriptsize{\textsf{SINR}}} > \Gamma]. 
\end{equation}
\begin{theorem}
Assuming that the random variables $r_z$ are i.i.d and Rayleigh distributed, it follows that the probability of coverage for the uplink cellular network is given by
\end{theorem}
\begin{equation}{\label{eq6}}
P_c({\scriptsize{\textsf{SINR}}},\Gamma) =  2\pi\lambda\int\limits_0^{+\infty} r e^{-\pi\lambda r^2-\mu\beta r^{\alpha(1-\epsilon)\sigma^2}}\mathscr{L}_{\mathcal{I}_z}\left(\mu\Gamma r^{\alpha(1-\epsilon)}\right)\text{d}r.
\end{equation}

where $\mathscr{L}_{\mathcal{I}_z}(\cdot)$ is the Laplace transform of the aggregate interference which is given by 
\begin{equation}{\label{eq7}}
\mathscr{L}_{\mathcal{I}_z}(s) = \exp\left(-2\pi\lambda \int\limits_r^{+\infty} \left(1-\mathbb{E}_{r_z}\left[\frac{\mu}{\mu + sr_z^{\alpha\epsilon}x^{-\alpha}}  \right]  \right)x \text{d}x \right). 
\end{equation}
\begin{proof}
 The proof of the Laplace Transform of the interference is provided in Appendix A.
\end{proof}
\subsection{Rate Analysis}
 The average uplink achievable rate, expressed in (nats/sec/Hz), is defined as the maximum error-free date rate  exchanged between a randomly selected user with its serving BS. It can be expressed as follows
 \begin{equation}\label{eq8}
\mathcal{I}({\scriptsize{\textsf{SINR}}}) = \mathbb{E}[\log(1+{\scriptsize{\textsf{SINR}}})].  
\end{equation}
\begin{theorem}
For the uplink cellular network, the achievable rate of the randomly selected user with its target BS is expressed as
\end{theorem}
\begin{equation}\label{eq9}
\begin{split}
\mathcal{I}({\scriptsize{\textsf{SINR}}}) =& 2\pi\lambda\int_{r>0}re^{-\pi\lambda r^2}\int_{x>0}e^{-\sigma^2\mu\frac{e^x-1}{r^{\alpha(\epsilon-1)}}}\mathscr{L}_{\mathcal{I}_z}\left(\mu\frac{e^x-1}{r^{\alpha(\epsilon-1)}}\right)\text{d}x\text{d}r.   
\end{split}
\end{equation}
\begin{proof}
 The proof is detailed in Appendix B.
\end{proof}

\section{Optical System Model}
In this Section, we provide the framework analysis of the optical sub-system connecting the BS with the data center. The FSO channel model consists of three components which are the atmospheric turbulences $I_a$, the pathloss $I_\ell$, and the pointing errors $I_p$. The aggregate FSO channel gain $I$ can be written as
\begin{equation}\label{eq16}
I = I_a l_\ell I_p.  
\end{equation}
\subsection{Atmospheric Turbulences}
We adopt the M\'alaga\footnote[1]{M\'alaga distribution is based on a physical model and it is widely employed as a fading model for the optical radio channel as it reflects a wide range of turbulences such as Lognormal, Gamma-Gamma, and Double-Weibull models, etc. More details about the derivation of this model and how to generate the samples are provided in \cite{malaga}} ($\mathcal{M}$-distribution) to model the atmospheric turbulences fading. In fact, this channel model consists of three components: the line-of-sight (LOS), the quasi-forward scattering, and energy scattered to the RX by off-axis eddies components, denoted by $(U_L)$, $(U_S^C)$, and $(U_S^G)$, respectively. The pdf of the turbulences-induced fading is given by
\begin{equation}\label{eq17}
f_{I_a}(I_a) = \Lambda\sum_{n=1}^{\kappa}\sigma_nI_a^{\frac{\nu+n}{2}-1}  K_{\nu-n}\left(2\sqrt{\frac{\nu\kappa I_a}{\zeta\kappa+\Omega^{\prime}}}\right)
\end{equation}
where $K_{\nu}(\cdot)$ is the $\nu$-th modified Bessel function of the second kind, $\zeta = \mathbb{E}[|U_S^C|^2 ] = 2b_0(1-\rho)$, $2b_0 = \mathbb{E}[|U_S^C|^2 + |U_S^G|^2]$ is the average power of the LOS path, $\Omega = \mathbb{E}[|U_L|^2]$ is the total scattered power, $\nu$ is a positive parameter related the large scale cells of the scattering process, $\kappa$ is a non-negative integer to account for the amount of fading, and $\Omega^{\prime} = \Omega + 2\rho b_0+2\sqrt{2\rho b_0\Omega}\cos(\theta_A - \theta_B)$ stands for the average power received from the coherent component. Note that $\rho$ (0$\leq\rho\leq$ 1) is the scattered power coupled with the LOS component, while $\theta_A$ and $\theta_B$ are the phases of the LOS and the coupled-to-LOS component. The parameters $\Lambda$ and $\sigma_n$ are given by
\begin{equation}\label{eq18}
\Lambda = \frac{2\nu^{\frac{\nu}{2}}}{\zeta^{1+\frac{1}{\nu}}\Gamma(\nu)}\left(\frac{\zeta\kappa}{\zeta\kappa+\Omega^{\prime}} \right)^{\kappa+\frac{\nu}{2}}   
\end{equation}
\begin{equation}\label{eq19}
\sigma_n = {\kappa-1 \choose n-1}\frac{(\zeta\kappa+\Omega^{\prime})^{1-\frac{n}{2}}}{(n-1)!}\left(\frac{\Omega^{\prime}}{\zeta} \right)^{n-1} \left(\frac{\nu}{\kappa} \right)^{\frac{n}{2}}  
\end{equation}
Note that $\Gamma(\cdot)$ is the incomplete upper Gamma function.
\subsection{Atmospheric Path Loss}
The path loss is deterministic and it is given by \cite[Eq.~(4)]{pathloss}
\begin{equation}\label{eq20}
    I_\ell = \frac{\pi a^2}{(\theta L_{\text{RD}})^2}\exp(-\sigma L_{\text{RD}})
\end{equation}
where $a$ is the radius of the receiver aperture, $\theta$ is the receive beam divergence, $L_{\text{RD}}$ is the optical link distance between the BS/relay and the data center, and $\sigma$ is the weather attenuation coefficient.

\subsection{Pointing Errors}
\begin{figure}[H]
    \centering
    \includegraphics[width=7cm]{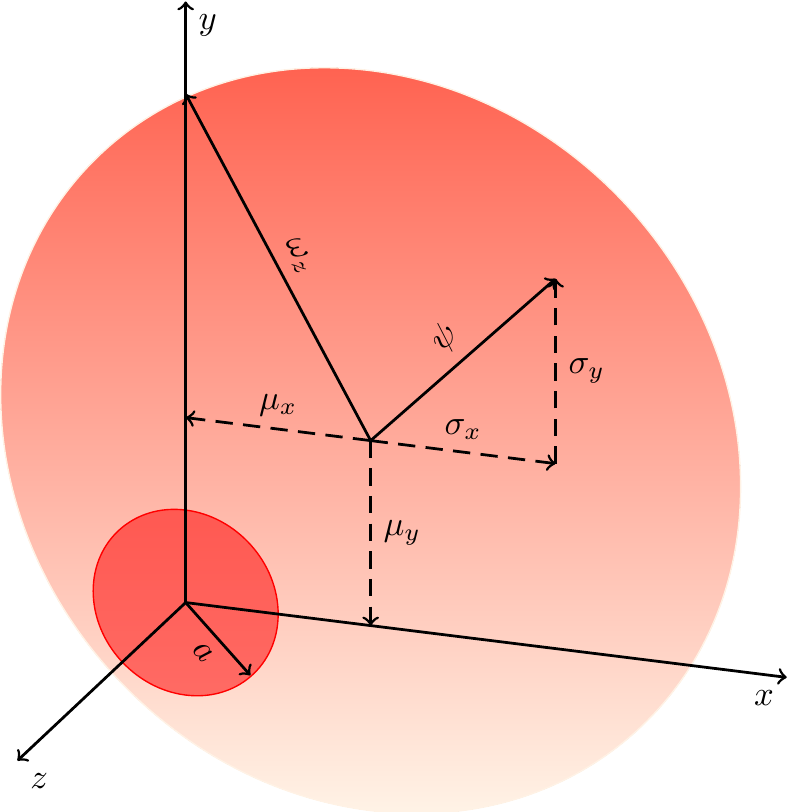}
    \caption{Beam footprint with generalized Beckmann pointing errors on the receiver aperture plane.}
    \label{fig3}
\end{figure}
During the radio propagation, the optical signal is subject to some fluctuations introduced by seismic activities or buildings swaying. This perturbation is translated into a misalignment between $R$ and $D$ or also called the pointing error which causes additional power losses. Thereby, the pointing error made by Jitter is modeled as follows \cite{aggregate}
\begin{equation}
    I_p(\psi;z) \cong A_0 \exp\left(-\frac{\psi^2}{\omega^2_{z_{eq}}}\right),~~\psi\geq 0
\end{equation}
where $r$ is the radial displacement at the receiver aperture, $\omega^2_{z_{eq}} = \frac{\sqrt{\pi} \text{erf}(v)}{2v\exp(-v^2)}\omega^2_z$ is the equivalent beam waist, $\omega_z$ is the beam waist at distance $z$, $A_0 = [\text{erf}(v)]^2$, $v = \sqrt{\frac{\pi}{2}}\frac{a}{\omega_z}$, $a$ is the radius of the receiver aperture, and $\text{erf}(\cdot)$ is the error function. Note that the beam waist $\omega_z$ is related to the Rytov variance $\sigma^2_{\text{R}}$ \cite[Eq.~(15)]{49} which is given by
\begin{align}
\sigma^2_{\text{R}} = 1.23 C_n^2 k^{7/6} L_{\text{RD}}^{11/6}   
\end{align}
where $C_n^2$ is the refractive index, $k$ is the wave number given by $k = 2\pi/\lambda$, and $\lambda$ is the wavelength.

The radial displacement $r$ can be expressed in the Cartesian coordinate as $\psi^2 = x^2+y^2$, where $x$, and $y$ are the horizontal and vertical displacements, respectively. Note that $x$, and $y$ are independent Gaussian random variables such as $x \backsim \mathcal{N}(\mu_x,\sigma_x)$, and $y \backsim \mathcal{N}(\mu_y,\sigma_y)$, respectively. Furthermore, it is noteworthy to define new parameters required for the analysis such as the ratios between the equivalent beam radius at the receiver and the corresponding pointing errors displacement standard deviation (Jitter) at the receiver such as $\phi_x = \omega_{z_{eq}}/2\sigma_x$, and $\phi_y = \omega_{z_{eq}}/2\sigma_y$. 

Since the radial displacement $\psi$ is distributed according to the Beckmann distribution, the generalized expression of the PDF is given by
\begin{equation}\label{psi}
    f_{\psi}(t) = \frac{t}{2\pi\sigma_x\sigma_y}\int\limits_0^{2\pi}\exp\left(-\frac{(t\cos(\theta)-\mu_x)^2}{2\sigma_x^2} - \frac{(t\sin(\theta)-\mu_y)^2}{2\sigma_y^2}  \right)d\theta,~~\psi \geq 0.
\end{equation}
Since the integral involved in (\ref{psi}) does not have a closed-form expression, the PDF of the displacement distribution is not mathematically tractable. Fortunately, the work in \cite{beckmann} had proposed an approach to approximate the Beckmann model to a modified Rayleigh distribution where the new PDF of the radial displacement can be expressed as 
\begin{equation}
f_{\psi}(t)\cong \frac{t}{\sigma^2_s}\exp\left(-\frac{t^2}{2\sigma^2_s} \right),~~t\geq 0.    
\end{equation}
Consequently, the new PDF of the pointing errors fading $I_p$ is given by
\begin{align}
    f_{I_p}(I_p) = \frac{g^2}{(A_0\eta)^{g^2}}I_p^{g^2-1},~0\leq I_p\leq A_0\eta
\end{align}
where $g^2 = \frac{\omega_{z_{eq}}}{2\sigma_s^2}$ is the pointing error coefficient, and $\sigma_s^2$ is the Jitter variance at the receiver. The parameters $\sigma_s^2$ and $\eta$ can be expressed as
\begin{equation}
\sigma_s^2 = \left(\frac{3\mu_x^2\sigma_x^4+3\mu_y^2\sigma_y^4+\sigma_x^6+\sigma_y^6}{2}\right)^{\frac{1}{3}}.    
\end{equation}
\begin{equation}
\eta = \exp\left(\frac{1}{\sigma_s^2}-\frac{1}{2\phi_x^2}-\frac{1}{2\phi_y^2}-\frac{\mu_x^2}{2\sigma_x^2\phi_x^2} - \frac{\mu_y^2}{2\sigma_y^2\phi_y^2} \right).    
\end{equation}
TABLES \ref{fso1} and \ref{weather1} summarize the main parameters values for the FSO system.
\begin{table}[H]
\renewcommand{\arraystretch}{.8}
\caption{FSO Sub-System}\label{fso1}
\centering
\begin{tabular}{|l|c|c|}
\hline\hline
\bfseries Parameter & \bfseries Symbol & \bfseries Value\\
\hline
Wavelength & $\lambda$& 1550 nm\\
\hline
Receiver Radius & $a$& 5 cm\\
\hline
Divergence Angle& $\theta$ & 10 mrad\\
\hline
Noise Variance& $\sigma^2_{\text{RD}}$& $\text{10}^{\text{-7}}$ A/Hz\\
\hline
Refractive Index& $C_n^2$ & $5\cdot10^{-14}$\\
\hline\hline
\end{tabular}
\end{table}
\begin{table}[H]
\renewcommand{\arraystretch}{.8}
\caption{Weather Dependent Parameters of FSO Channel \cite{mmwavej,mmwavec}}
\label{weather1}
\centering
\begin{tabular}{|l|c|c|}
\hline\hline
\bfseries Weather Conditions & \bfseries $\sigma$ (dB/km) &  \bfseries $C^2_n$\\
\hline
Clear Air & 0.43  & 5$\cdot\text{10}^{\text{-14}}$\\
\hline
Moderate Fog & 42.2  & 2$\cdot\text{10}^{\text{-15}}$\\
\hline
Moderate Rain (12.5 mm/h) & 5.8  & 5$\cdot\text{10}^{\text{-15}}$\\
\hline\hline
\end{tabular}
\end{table}

\subsection{FSO System Performance Analysis}
\subsubsection{SNR Statistics}
In case of the heterodyne detection, the average SNR $\mu_1$ is given by $\mu_1 = \frac{\e{I}}{\sigma^2_{\text{RD}}}$. Regarding the IM/DD detection, the average electrical SNR $\mu_2$ is given by $\mu_2 = \frac{\e{I}^2}{\sigma^2_{\text{RD}}}$ while the instantaneous optical SNR is $\gamma_{r} = \frac{I^2}{\sigma^2_{\text{RD}}}$. Unifying the two detection schemes yields $\gamma_{r} = \frac{I^r}{\sigma_{\text{RD}}^2}$. Note that $r$ can take the values 1 and 2 for the heterodyne and IM/DD, respectively.

The average SNR $\overline{\gamma}_r$\footnote[2]{The average SNR $\overline{\gamma}_r$ is defined as $\overline{\gamma}_r = \e{I^r}/\sigma_{\text{RD}}^2$, while the average electrical SNR $\mu_r$ is given by $\mu_r = \e{I}^r/\sigma_{\text{RD}}^2$. Therefore, the relation between the average SNR and the average electrical SNR is trivial given that $ \frac{\e{I^2}}{\e{I}^2} = \sigma^2_{\text{si}} + 1$, where $\sigma^2_{\text{si}}$ is the scintillation index \cite{scin}.} can be expressed as
\begin{align}
 \overline{\gamma}_r = \frac{\e{I^r}}{\e{I}^r}\mu_r   
\end{align}
where $\mu_r$ is the average electrical SNR given by
\begin{equation}
    \mu_r = \frac{\e{I^r}}{\sigma_{\text{RD}}^2}.
\end{equation}
where $G_{p,q}^{m,n}(\cdot)$ is the Meijer-G function, $\tau_n = \sigma_n\left(\frac{g\kappa+\Omega^{\prime}}{\nu\kappa} \right)^{\frac{\nu +n}{2}}$, and $\delta = \frac{g^2\nu\kappa(g+\Omega^{\prime})}{(g^2+1)(g\kappa+\Omega^{\prime})}$. The CDF of the instantaneous SNR is obtained by
\begin{equation}\label{fsocdf}
F_{\gamma_r}(x) = \Lambda^{\prime}\sum_{n=1}^{\kappa}\varrho_n G_{r+1,3r+1}^{3r,1}\Bigg(\frac{\delta^\prime}{\mu_r}x  ~\bigg|~\begin{matrix} 1,~\Delta(r:g^2+1) \\ \Delta(r:g^2),~\Delta(r:\kappa) \end{matrix} \Bigg)
\end{equation}
where $\Lambda^{\prime} = \frac{g^2\Lambda}{2^r(2\pi)^{2r-1}}$, $\delta^{\prime} = \frac{\delta^r}{r^{2r}}$, $\varrho_n = \tau_n r^{\nu+n-1}$, and $\Delta(n:x) = \frac{x}{n},\ldots,\frac{x+n-1}{n}$.
From the PDF expression, we can derive the analytical $n$-th moment as follows
\begin{equation}\label{moment}
\mathbb{E}[\gamma_r^n] =  \frac{rg^2\Lambda\Gamma(nr+\nu)}{2^r(nr+g^2)\delta^{nr}}\sum_{m=1}^\kappa \tau_m \Gamma(nr+m)\mu_r^n.   
\end{equation}
\subsubsection{Achievable Rate}
For IM/DD and direct detections, the ergodic achievable rate of the FSO link can be written as
\begin{equation}
\mathcal{I}({\scriptsize{\textsf{SNR}}},\varpi) = \mathbb{E}[\log(1+\varpi{\scriptsize{\textsf{SNR}}})] = \int\limits_0^{\infty}\log(1+\varpi x) f_{\gamma_r}(x)\text{d}x.     
\end{equation}
After converting $\log(\cdot)$ into Meijer-G function \cite[Eq.~(07.34.03.0456.01)]{wolfram} and using the identity \cite[Eq.~(2.24.1.1)]{brysh}, the closed-form expression of the rate achieved by the FSO link is given by
\begin{equation}\label{ratefso}
\begin{split}
\mathcal{I}({\scriptsize{\textsf{SNR}}},\varpi) =& \Lambda^{\prime}\sum_{n=1}^\kappa \varrho_n G_{r+2,3r+2}^{3r+2,1}\Bigg(\frac{\delta^\prime}{\varpi\mu_r}  ~\bigg|~\begin{matrix} 0,~1,~\Delta(r:g^2+1) \\ \Delta(r:g^2),~\Delta(r:\nu),~\Delta(r:n) \end{matrix} \Bigg),    
\end{split}
\end{equation}
where $\varpi$ can take 1 or $e/2\pi$ for heterodyne and IM/DD detections, respectively. 
\begin{corollary}
After considering the following approximation for lower $x$, ($\log(x+1) \cong x$), the rate can be approximated at low {\scriptsize{\textsf{SNR}}} ($\mu_r << 1$) as follows
\end{corollary}
\begin{equation}\label{low}
\mathcal{I}_{\sf{low}}({\scriptsize{\textsf{SNR}}},\varpi) \cong \varpi\mathbb{E}[\gamma_r^{n=1}] = \frac{\varpi rg^2\Lambda\Gamma(r+\nu)}{2^r(g^2+r)\delta^r}\sum_{m=1}^\kappa \tau_m\Gamma(m+r).   
\end{equation}

\begin{corollary}
The high {\scriptsize{\textsf{SNR}}} ($\mu_r >> 1$)  approximation of the achievable rate can be derived by using \cite[Eq.~(07.34.06.0001.01)]{wolfram} to expand the Meijer-G function in (\ref{ratefso}) as follows
\end{corollary}
\begin{equation}\label{high1}
\mathcal{I}_{\sf{high}}({\scriptsize{\textsf{SNR}}},\varpi) \cong \Lambda \sum_{m=1}^\kappa\sum_{n=1}^{3r+2}\varrho_m\left(\frac{\varpi\mu_r}{\Delta^\prime}\right)^{-\vartheta_{2,n}}\frac{\Gamma(1+\vartheta_{2,n})\prod_{\ell=1;\ell\neq n}^{3r+2}\Gamma(\vartheta_{2,\ell}-\vartheta_{2,n})}{\Gamma(1-\vartheta_{2,n})\prod_{\ell=1}^{r+2}\Gamma(\vartheta_{1,\ell}-\vartheta_{2,n})}  
\end{equation}
where $\vartheta_1 = \Delta(g^2+1:r)$, and $\vartheta_2 = [\Delta(r:g^2),~\Delta(r:\nu),~\Delta(r:m)$].
We can also characterize the asymptotic high SNR ($\mu_r >> 1$) of the achievable rate by considering an approach based on the $n$th moment stated by the following theorem. 
\begin{corollary}
At high {\scriptsize{\textsf{SNR}}} ($\mu_r >> 1$), the achievable rate can be approximated by
\end{corollary}
\begin{equation}\label{high2}
\begin{split}
&\mathcal{I}_{\sf{high}}({\scriptsize{\textsf{SNR}}},\varpi) \cong  \frac{\partial \mathbb{E}[\gamma_r^n]}{\partial n}\Bigr\rvert_{n = 0} = \frac{r\Lambda\Gamma(\nu)}{2^r}\sum_{m=1}^\kappa
\tau_m\Gamma(m)\left(r \left[  -\frac{1}{g^2}-\log(\delta)+\psi(\nu)+\psi(m)\right]+\log(\varpi \mu_r)\right).    
\end{split}
\end{equation}
where $\psi(\cdot)$ is the Digamma function.
\begin{corollary}
The achievable rate can be further characterized by deriving an upper bound using Jensen's inequality as follows
\end{corollary}
\begin{equation}
\mathcal{I}_{\sf{ub}}({\scriptsize{\textsf{SNR}}},\varpi) = \log(1 + \varpi\mathbb{E}[\gamma_r]).    
\end{equation}

\section{Hybrid Networking System Analysis}
In this Section, we present the performance analysis of the hybrid cellular and FSO network in terms of the coverage probability, and the ergodic achievable rate.
\subsection{Repetition-Coded DF Relaying}
The choice of the DF relaying scheme is relevant as the cellular network and the FSO system are independent. Thereby, the overall system analysis can follow up from the fact that the system can be described as a dual-hop relaying network. The relations between the results of the two sub-systems will be analyzed in the following Subsections.
\subsubsection{SINR Analysis}
Since the BS behaves as a DF relay, the equivalent ${\scriptsize{\textsf{SINR}}}$ can be written as
\begin{equation}
{\scriptsize{\textsf{SINR}}} = \min({\scriptsize{\textsf{SINR}}}_{\sf{uplink}},{\scriptsize{\textsf{SNR}}}_{\sf{backhaul}}).    
\end{equation}
\subsubsection{Coverage Probability}
Assuming the independence between the uplink and backhaul CSIs, It follows that the coverage probability of the hybrid system is expressed as
\begin{equation}
P_c({\scriptsize{\textsf{SINR}}},\Gamma) = \mathbb{P}[{\scriptsize{\textsf{SINR}}}_{\sf{uplink}} > \Gamma] \mathbb{P}[{\scriptsize{\textsf{SNR}}}_{\sf{backhaul}}> \Gamma].   
\end{equation}
\subsubsection{Achievable Rate}
Since DF relaying is assumed, the ergodic achievable rate of the hybrid system is the minimum between the uplink and backhaul rates. It can be expressed as follows
\begin{equation}
\mathcal{I}({\scriptsize{\textsf{SINR}}},\varpi)= \min(\mathcal{I}({\scriptsize{\textsf{SINR}}}),\mathcal{I}({\scriptsize{\textsf{SNR}}},\varpi)).    
\end{equation}
\subsubsection{Diversity Gain}
For the uplink cellular network, the channels are Rayleigh distributed and both the UEs and BSs are equipped with a single antenna. Thereby, the diversity is at most one. After using \cite[Eq.~(07.34.06.0001.01)]{wolfram} to expand the Meijer-G function involved in the cdf (\ref{fsocdf}) at high SNR, the diversity gain achieved by the SISO FSO link is $\min\left(\frac{g^2}{r},~\frac{\nu}{r},~\frac{\kappa}{r}\right)$.
Consequently, the diversity gain ${\textsf{G}}_{\sf{d}}$ achieved by the hybrid networking system can be obtained by
\begin{equation}\label{diversity}
{\textsf{G}}_{\sf{d}} = \min\left(1,~\min\left(\frac{g^2}{r},~\frac{\nu}{r},~\frac{\kappa}{r} \right)\right).    
\end{equation}
\begin{remark}
An interesting conclusion drawn from (\ref{diversity}) states that the diversity gain depends to a large extent on the state of the optical channel as it is affected by the pointing errors, the atmospheric turbulences, and the optical detection schemes.
\end{remark}
\subsection{Intelligent Reflecting Surfaces}
\begin{figure}[H]
\begin{subfigure}[b]{0.5\textwidth}
\centering
\includegraphics[width=\linewidth]{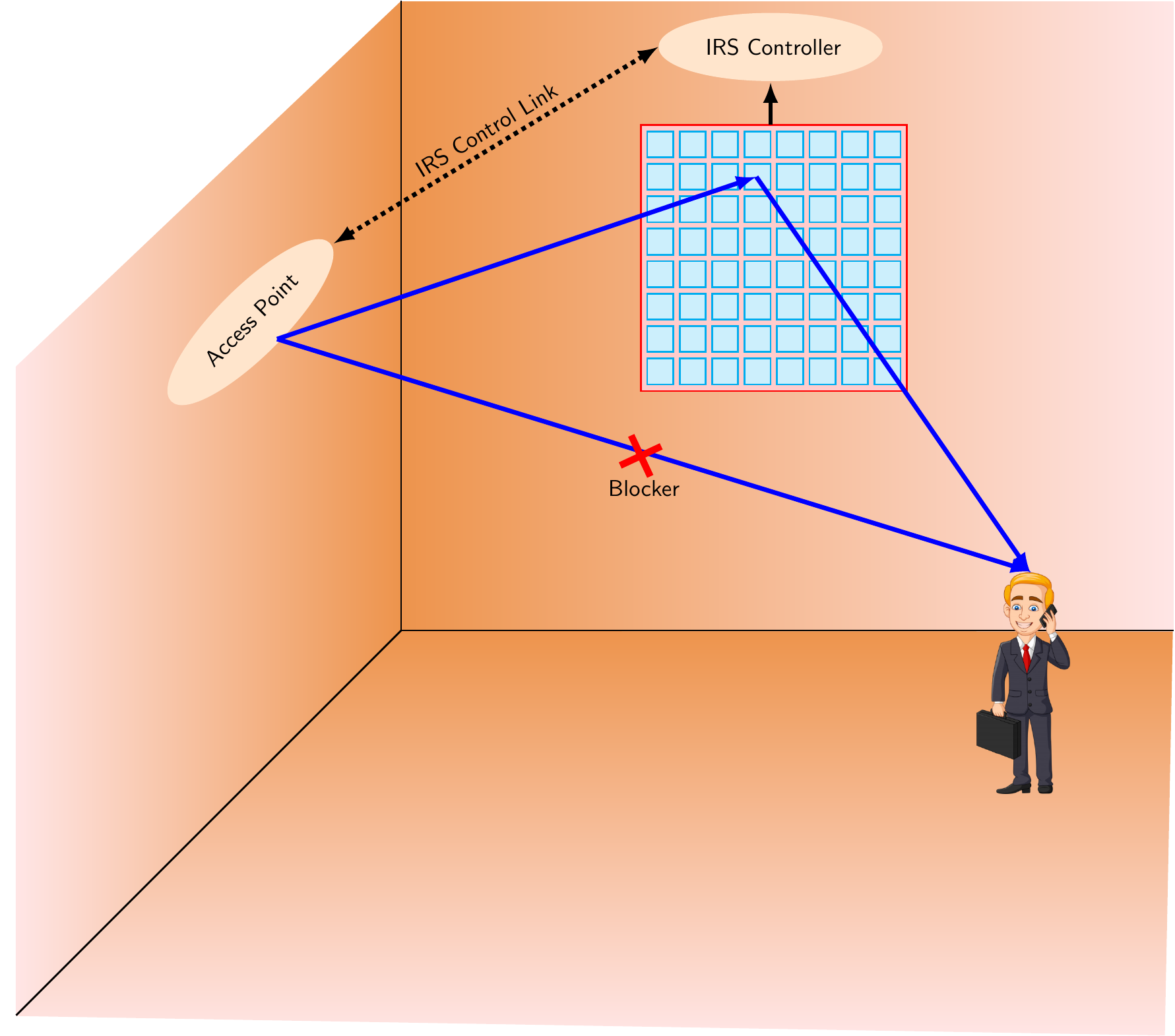}
    \caption{Indoor.}
    \label{a1}
    \end{subfigure}
    \begin{subfigure}[b]{0.5\textwidth}
\centering
\includegraphics[width=\linewidth]{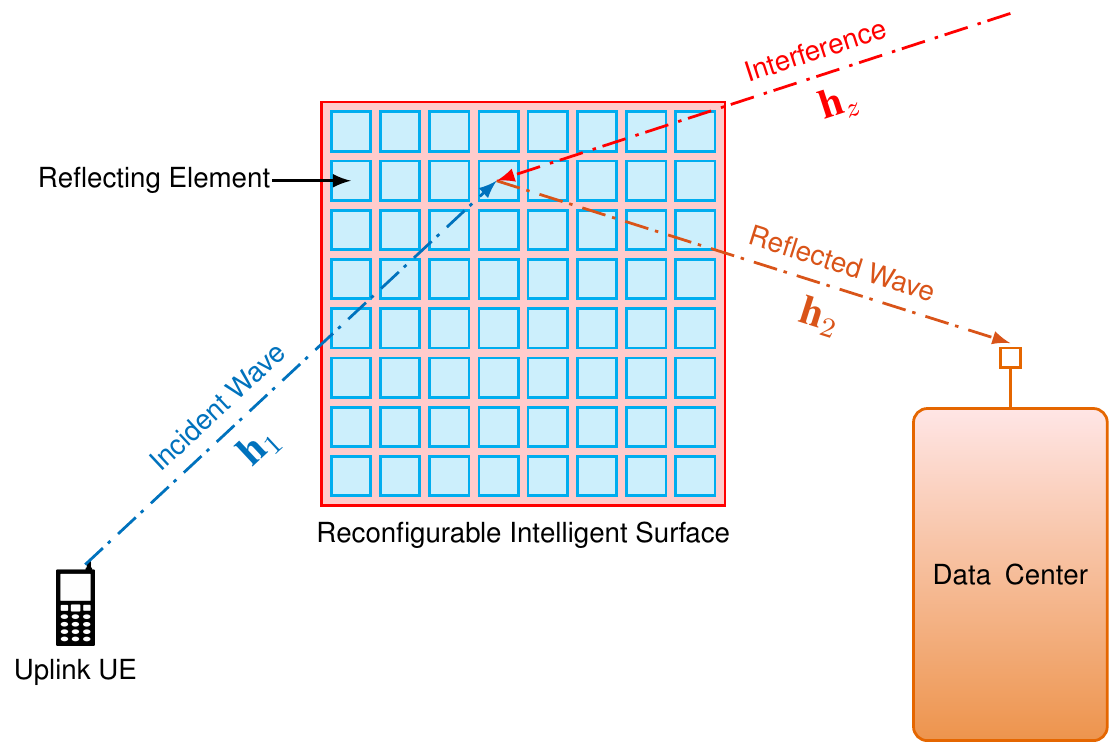}
    \caption{Outdoor.}
    \label{b1}
    \end{subfigure}
    \caption[map]{Intelligent reflecting surfaces with $N$ reflecting elements. Each element is characterized by a phase shifter to control the direction of the reflected wave. Such surface can be interpreted as a massive squared array panel of 8 x 8 elements. }
    \label{irs}
\end{figure}
Fig.~\ref{irs} illustrates a communication scenario between the uplink UE and the data center through an IRS. Such panel is attached to a surrounding building's fascade, and the transmit signal propagates to the data center via the assistance of IRS that is capable to reconfigure and control the direction of the reflected wave. In this framework analysis, we consider an IRS panel with $N$ discrete elements wherein each element is characterized by a phase shifter. Therefore, the IRS is fully characterized by the phases shifters of all the elements. An appropriate mathematical representation of the IRS elements is given by
\begin{equation}
\boldsymbol{\Phi} = \text{diag}\left(e^{j\phi_1},\ldots, e^{j\phi_N}  \right).    
\end{equation}
\subsubsection{SINR Analysis}
For the IRS scenario, the {\scriptsize{\textsf{SINR}}} can be expressed as
\begin{equation}\label{sinr-irs}
    {\scriptsize{\textsf{SINR}}} = \frac{|\textbf{h}_1^*\boldsymbol{\Phi} \textbf{h}_2|^2}{\sum_{z\in\mathcal{Z}}r_z^{\alpha\epsilon}d_z^{-\alpha}|\textbf{h}^*_z\boldsymbol{\Phi} \textbf{h}_2|^2 + \sigma^2}.
\end{equation}

\subsubsection{Optimization}
According to (\ref{sinr-irs}), the performance is highly depending on the phases shifters of the IRS. Then, it is required to optimize the phases shifters to maximize the {\scriptsize{\textsf{SINR}}} and the spectral efficiency. Note that it is straightforward that maximizing the spectral efficiency is equivalent to maximizing the {\scriptsize{\textsf{SINR}}}. For this purpose, we consider the {\scriptsize{\textsf{SINR}}} as the objective function and we formulate the optimization problem $\boldsymbol{P}$ as follows
\begin{equation}\label{opt}
\begin{split}
\boldsymbol{P}: \max\limits_{\boldsymbol{\Phi}} &~|\textbf{h}_1^*\boldsymbol{\Phi} \textbf{h}_2|  
\end{split}
\end{equation}
\begin{equation}\label{zf}
\text{subject to}~|\textbf{h}_z^*\boldsymbol{\Phi} \textbf{h}_2| = 0.
\end{equation}
The problem $\boldsymbol{P}$ is equivalent to maximizing the {\scriptsize{\textsf{SINR}}}, since the objective function is to maximize the beamformed power (\ref{opt}) and minimize the interference (\ref{zf}). In other term, problem $\boldsymbol{P}$ is equivalent to find the optimal phases shifters of IRS that jointly maximize the beamformed power and minimize the interference. Note that (\ref{zf}) is also called the Zero-Forcing constraint. The problem $\boldsymbol{P}$ can be transformed into
\begin{equation}\label{opt1}
\begin{split}
\boldsymbol{P}: \max\limits_{\boldsymbol{\Phi}} &~|\boldsymbol{x}^*\textbf{h}_1|^2,~\text{subject to}~\boldsymbol{x}^*\textbf{h}_z = 0   
\end{split}
\end{equation}
where $\boldsymbol{x} = \boldsymbol{\Phi}\textbf{h}_2$. The solution $\boldsymbol{\hat{x}}$ to the problem (\ref{opt1}) is given by
\begin{equation}
\boldsymbol{\hat{x}} = \left(\boldsymbol{I} - \frac{\textbf{h}_z\textbf{h}^*_z}{\|\textbf{h}_z\|^2} \right)\textbf{h}_1 
\end{equation}
Therefore, the optimal phase shifter of the $n$-th reflecting element is expressed as
\begin{equation}
\phi_n = \text{angle}\left(\ddfrac{\left[\left(\boldsymbol{I} - \frac{\textbf{h}_z\textbf{h}^*_z}{\|\textbf{h}_z\|^2} \right)\textbf{h}_1\right]_n}{[\textbf{h}_2]_n} \right),~n = 1,\ldots, N. 
\end{equation}
\textit{\underline{Special case}:} For interference-free scenario ($\textbf{h}_z = 0$), the $n$-th optimal phase shifter of the IRS panel is given by
\begin{equation}
    \phi_n = \text{angle}\left( \frac{[\textbf{h}_1]_n}{[\textbf{h}_2]_n} \right) = - \text{angle}\left([\textbf{h}_1]_n[\textbf{h}_2]_n  \right),~n = 1,\ldots, N. 
\end{equation}
Under this scenario, we retrieve the same results derived by \cite{emil}.

\section{Numerical Results and Discussion}
In this section, we will present the numerical results of the analytical expressions of the system performance following their discussions along with some comparisons. Unless otherwise stated, TABLE \ref{param} summarizes the simulation parameter.
\begin{table}[H]
\renewcommand{\arraystretch}{.8}
\caption{Cellular System Parameters}\label{param}
\centering
\begin{tabular}{|l|c|c|}
\hline\hline
\bfseries Parameter & \bfseries Symbol & \bfseries Value\\
\hline
UE density & $\lambda$& 0.25 UE/km$^2$\\
\hline
Pathloss exponent & $\alpha$& 3.5\\
\hline
Uplink transmit power& $\mu^{-1}$ & 150 mW\\
\hline
Fractional power control & $\epsilon$&  0.6\\
\hline
Bandwidth & BW &  300 MHz\\
\hline
Noise power density& $\sigma^2$ &  173.8 - 10 $\log_{10}$(BW)\\
\hline\hline
\end{tabular}
\end{table}
\begin{figure}[H]
\centering
\setlength\fheight{5.5cm}
\setlength\fwidth{7.5cm}
%
%

\definecolor{mycolor1}{rgb}{0.00000,0.44700,0.74100}%
\definecolor{mycolor2}{rgb}{0.85000,0.33000,0.10000}%
\begin{tikzpicture}

\begin{axis}[%
width=0.951\fwidth,
height=\fheight,
at={(0\fwidth,0\fheight)},
scale only axis,
xmin=-20,
xmax=20,
xlabel style={font=\color{white!15!black}},
xlabel={\sffamily{Average SNR (dB)}},
ymin=0,
ymax=4,
ylabel style={font=\color{white!15!black}},
ylabel={\sffamily{Spectral Efficiency (nats/sec/Hz)}},
axis background/.style={fill=white},
axis x line*=bottom,
axis y line*=left,
axis line style = thick,
every major tick/.append style={thick, black},
every minor tick/.append style={thick, black},
legend style={at={(0.03,0.97)}, anchor=north west, legend cell align=left, align=left, draw=white!15!black,draw=none}
]
\addplot [color=black, line width=1.3pt]
  table[row sep=crcr]{%
-20	0.00988534981262924\\
-19	0.0124087390868738\\
-18	0.0155651595718446\\
-17	0.0195073791742827\\
-16	0.0244218751976058\\
-15	0.0305347059119483\\
-14	0.0381176696299573\\
-13	0.0474944090743281\\
-12	0.0590459632755937\\
-11	0.0732151076989215\\
-10	0.0905086832517972\\
-9	0.111497033618975\\
-8	0.136809689867558\\
-7	0.167126598097623\\
-6	0.20316449891079\\
-5	0.24565852578583\\
-4	0.29533964612827\\
-3	0.352909143776695\\
-2	0.419011835879272\\
-1	0.494210033900129\\
0	0.578960328480351\\
1	0.673595076546085\\
2	0.778310023997854\\
3	0.893158881307303\\
4	1.01805498255209\\
5	1.15277950539333\\
6	1.29699519781314\\
7	1.45026420318842\\
8	1.61206841793423\\
9	1.781830841768\\
10	1.95893655204038\\
11	2.14275220081679\\
12	2.33264324551412\\
13	2.52798843683771\\
14	2.72819136893782\\
15	2.93268912629422\\
16	3.14095823163414\\
17	3.35251821005679\\
18	3.56693314338043\\
19	3.78381160566434\\
20	4.00280535695077\\
};
\addlegendentry{\sffamily{Exact}}

\addplot [color=mycolor2, dash pattern={on 10pt off 1pt on 1pt off 1pt} , line width=1.3pt]
  table[row sep=crcr]{%
-20	-5.26314651407206\\
-19	-5.03288800477265\\
-18	-4.80262949547325\\
-17	-4.57237098617384\\
-16	-4.34211247687444\\
-15	-4.11185396757503\\
-14	-3.88159545827563\\
-13	-3.65133694897622\\
-12	-3.42107843967682\\
-11	-3.19081993037741\\
-10	-2.96056142107801\\
-9	-2.7303029117786\\
-8	-2.5000444024792\\
-7	-2.2697858931798\\
-6	-2.03952738388039\\
-5	-1.80926887458099\\
-4	-1.57901036528158\\
-3	-1.34875185598218\\
-2	-1.11849334668277\\
-1	-0.888234837383368\\
0	-0.657976328083963\\
1	-0.427717818784558\\
2	-0.197459309485153\\
3	0.032799199814251\\
4	0.263057709113656\\
5	0.493316218413061\\
6	0.723574727712465\\
7	0.953833237011869\\
8	1.18409174631127\\
9	1.41435025561068\\
10	1.64460876491008\\
11	1.87486727420949\\
12	2.10512578350889\\
13	2.3353842928083\\
14	2.5656428021077\\
15	2.79590131140711\\
16	3.02615982070651\\
17	3.25641833000592\\
18	3.48667683930532\\
19	3.71693534860473\\
20	3.94719385790413\\
};
\addlegendentry{\sffamily{High SNR}}

\addplot [color=black, dotted, line width=1.3pt]
  table[row sep=crcr]{%
-20	0.01\\
-19	0.0125892541179417\\
-18	0.0158489319246111\\
-17	0.0199526231496888\\
-16	0.0251188643150958\\
-15	0.0316227766016838\\
-14	0.0398107170553497\\
-13	0.0501187233627272\\
-12	0.0630957344480193\\
-11	0.0794328234724281\\
-10	0.1\\
-9	0.125892541179417\\
-8	0.158489319246111\\
-7	0.199526231496888\\
-6	0.251188643150958\\
-5	0.316227766016838\\
-4	0.398107170553497\\
-3	0.501187233627272\\
-2	0.630957344480193\\
-1	0.794328234724281\\
0	1\\
1	1.25892541179417\\
2	1.58489319246111\\
3	1.99526231496888\\
4	2.51188643150958\\
5	3.16227766016838\\
6	3.98107170553497\\
7	5.01187233627272\\
8	6.30957344480193\\
9	7.94328234724282\\
10	10\\
11	12.5892541179417\\
12	15.8489319246111\\
13	19.9526231496888\\
14	25.1188643150958\\
15	31.6227766016838\\
16	39.8107170553497\\
17	50.1187233627272\\
18	63.0957344480193\\
19	79.4328234724282\\
20	100\\
};
\addlegendentry{\sffamily{Low SNR}}

\addplot [color=mycolor1, dash pattern={on 10pt off 1pt on 0pt off 0pt}, line width=1.3pt]
  table[row sep=crcr]{%
-20	0.00995033085316809\\
-19	0.0125106683286301\\
-18	0.0157246490503978\\
-17	0.0197561783249809\\
-16	0.0248085710523884\\
-15	0.0311330736895024\\
-14	0.0390386937573615\\
-13	0.0489032276467801\\
-12	0.0611851559330244\\
-11	0.0764357397380491\\
-10	0.0953101798043249\\
-9	0.118576091042074\\
-8	0.147116845379613\\
-7	0.181926671751191\\
-6	0.224094014002177\\
-5	0.274769892408345\\
-4	0.335119300782909\\
-3	0.406256284130706\\
-2	0.489167170311102\\
-1	0.584630709400607\\
0	0.693147180559945\\
1	0.814889218700012\\
2	0.949684188909911\\
3	1.09703181202892\\
4	1.25615333798053\\
5	1.42606243890537\\
6	1.6056450697986\\
7	1.79373623684702\\
8	1.98918491977485\\
9	2.19090267473671\\
10	2.39789527279837\\
11	2.6092793420315\\
12	2.82428726752588\\
13	3.04226384853904\\
14	3.26265782394903\\
15	3.48501071318057\\
16	3.70894471984286\\
17	3.93415083641486\\
18	4.16037781643968\\
19	4.38742234501732\\
20	4.61512051684126\\
};
\addlegendentry{\sffamily{Upper Bound}}

\end{axis}
\end{tikzpicture}%
    \caption{Achievable rate of the FSO link under heterodyne detection: Exact closed form, upper bound, low and high SNR approximations are presented.}
     \label{fig4}
\end{figure}
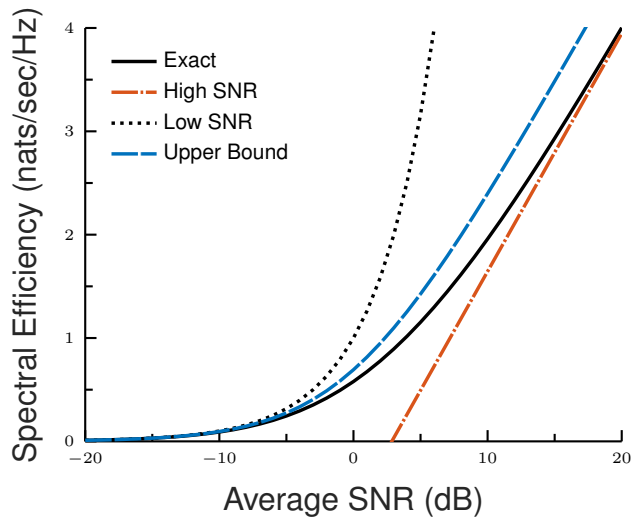
Fig.~\ref{fig4} illustrates the characteristics of the capacity in terms of exact, low and high SNR regimes, and upper bound analytical expressions. For low SNR regime, the characteristic (dotted) is perfectly tight to the exact analytical expression, however, it pronouncedly deviates around 0 dB. In addition, the high SNR expansion perfectly converges to the exact expression up to 15 dB. Furthermore, the Jensen's upper bound presents a perfect tightness to the exact capacity at low SNR, however, a gap between the two characteristics appears around 5 dB and continues to enlarge with the average SNR.
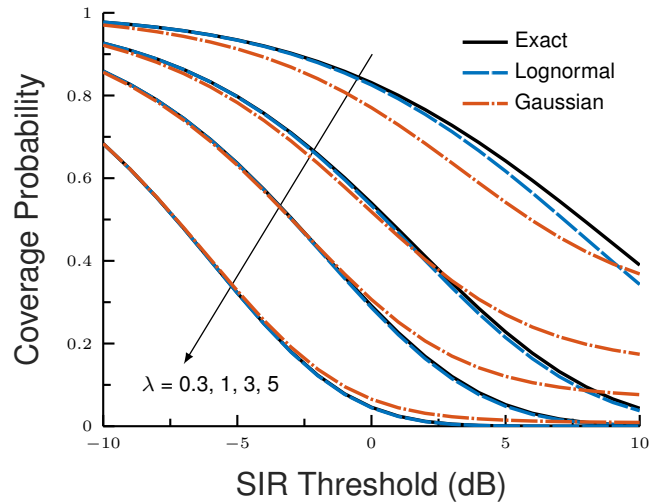
\begin{figure}[H]
\centering
\setlength\fheight{5.5cm}
\setlength\fwidth{7.5cm}
%
%
\definecolor{mycolor1}{rgb}{0.00000,0.44700,0.74100}%
\definecolor{mycolor2}{rgb}{0.85000,0.33000,0.10000}%
\usetikzlibrary{positioning}
\begin{tikzpicture}

\begin{axis}[%
width=0.951\fwidth,
height=\fheight,
at={(0\fwidth,0\fheight)},
scale only axis,
xmin=-10,
xmax=10,
xlabel style={font=\color{white!15!black}},
xlabel={\sffamily{SIR Threshold (dB)}},
ymin=0,
ymax=1,
ylabel style={font=\color{white!15!black}},
ylabel={\sffamily{Coverage Probability}},
axis background/.style={fill=white},
legend style={legend cell align=left, align=left,
axis x line*=bottom,
axis y line*=left,
axis line style = thick,
every major tick/.append style={thick, black},
every minor tick/.append style={thick, black},
draw=white!15!black,draw=white!15!black,draw=none}
]

\draw [-latex,line width=.5pt] (0.03,0.9) to (-7,0.15);
\node[] at (-6,0.1){\scriptsize\sffamily{$\lambda$ = 0.3,~1,~3,~5}};

\addplot [color=black, line width=1.3pt]
  table[row sep=crcr]{%
-10	0.977437872503421\\
-9	0.971898135077078\\
-8	0.965085076603275\\
-7	0.956749547878386\\
-6	0.946613876553315\\
-5	0.934377366037894\\
-4	0.919725708592224\\
-3	0.902344551231407\\
-2	0.881936784124606\\
-1	0.858242326320615\\
0	0.831058505794232\\
1	0.800258855947234\\
2	0.765808457514054\\
3	0.727774777085272\\
4	0.686333967727067\\
5	0.641773374393455\\
6	0.594491216437171\\
7	0.544994053796109\\
8	0.493891881237146\\
9	0.441889851320066\\
10	0.389774995925682\\
};
\addlegendentry{\sffamily{Exact}}

\addplot [color=mycolor1, dash pattern={on 10pt off 1pt on 0pt off 0pt} , line width=1.3pt]
  table[row sep=crcr]{%
-10	0.977403364610335\\
-9	0.971835275319685\\
-8	0.964972290106069\\
-7	0.956550583464291\\
-6	0.946269451648834\\
-5	0.933793440705133\\
-4	0.918758100613856\\
-3	0.900780427930031\\
-2	0.879474937717205\\
-1	0.854475958571337\\
0	0.825466121444716\\
1	0.792210102021086\\
2	0.754591529940877\\
3	0.712649732265177\\
4	0.666611856304431\\
5	0.616915206072021\\
6	0.564214626339098\\
7	0.509370709735171\\
8	0.453416559715061\\
9	0.397503661361398\\
10	0.342830687716585\\
};
\addlegendentry{\sffamily{Lognormal}}

\addplot [color=mycolor2, dash pattern={on 10pt off 1pt on 1pt off 1pt} , line width=1.3pt]
  table[row sep=crcr]{%
-10	0.970595966247543\\
-9	0.963267802488625\\
-8	0.954202614732242\\
-7	0.943038333823349\\
-6	0.929364706666051\\
-5	0.912732318982081\\
-4	0.892672148481804\\
-3	0.868729786128104\\
-2	0.840518189372153\\
-1	0.807790847172945\\
0	0.770532463013445\\
1	0.729055926309679\\
2	0.684083077165005\\
3	0.63677615559549\\
4	0.588684294741485\\
5	0.541584201039306\\
6	0.497230249128897\\
7	0.457075698990322\\
8	0.422055845938632\\
9	0.392507899781026\\
10	0.368242117770846\\
};
\addlegendentry{\sffamily{Gaussian}}

\addplot [color=black, line width=1.3pt, forget plot]
  table[row sep=crcr]{%
-10	0.926752732068122\\
-9	0.909359962910717\\
-8	0.888284328500663\\
-7	0.862967060302252\\
-6	0.832868087654\\
-5	0.797518988980504\\
-4	0.756590854798951\\
-3	0.709972184332045\\
-2	0.657846935037883\\
-1	0.600758277487502\\
0	0.539641485816297\\
1	0.475811531337272\\
2	0.410897842715636\\
3	0.34672886640453\\
4	0.285179505154795\\
5	0.228002172097978\\
6	0.176665389173764\\
7	0.132222656890079\\
8	0.0952298186162352\\
9	0.0657224015894855\\
10	0.0432560228100782\\
};
\addplot [color=mycolor1, dash pattern={on 10pt off 1pt on 0pt off 0pt} , line width=1.3pt, forget plot]
  table[row sep=crcr]{%
-10	0.926706213549461\\
-9	0.90927467663893\\
-8	0.888130724646989\\
-7	0.862696109743125\\
-6	0.832401511692054\\
-5	0.796737394239778\\
-4	0.755321800712407\\
-3	0.707982461209813\\
-2	0.654845748874919\\
-1	0.596419276858246\\
0	0.533649694378869\\
1	0.467934712492069\\
2	0.40107100180993\\
3	0.335128991958741\\
4	0.27226134099647\\
5	0.214470599783532\\
6	0.163377464224206\\
7	0.120037398280378\\
8	0.084846016165847\\
9	0.0575530428281453\\
10	0.037377094934578\\
};
\addplot [color=mycolor2, dash pattern={on 10pt off 1pt on 1pt off 1pt} , line width=1.3pt, forget plot]
  table[row sep=crcr]{%
-10	0.921610689685261\\
-9	0.902920395724337\\
-8	0.880256157927013\\
-7	0.853027484164772\\
-6	0.820688554282126\\
-5	0.782817936640713\\
-4	0.739225443705752\\
-3	0.690079177248617\\
-2	0.636032702068472\\
-1	0.578315739830933\\
0	0.518738329792488\\
1	0.459560293952169\\
2	0.403208536990392\\
3	0.351886416835704\\
4	0.307188335798929\\
5	0.269860664676271\\
6	0.239799003657484\\
7	0.216260321997478\\
8	0.198172881546389\\
9	0.184412897304312\\
10	0.173976608480983\\
};
\addplot [color=black, line width=1.3pt, forget plot]
  table[row sep=crcr]{%
-10	0.858870626395729\\
-9	0.826935542144981\\
-8	0.789049048259874\\
-7	0.744712147166712\\
-6	0.693669251432432\\
-5	0.636036537784486\\
-4	0.572429721565407\\
-3	0.504060502525216\\
-2	0.432762589938737\\
-1	0.360910507969751\\
0	0.291212933214021\\
1	0.22639661335352\\
2	0.168837037148364\\
3	0.120220906798171\\
4	0.0813273501603338\\
5	0.0519849904813959\\
6	0.0312106597319176\\
7	0.0174828309950716\\
8	0.00906871835368105\\
9	0.0043194340706896\\
10	0.00187108350934601\\
};
\addplot [color=mycolor1, dash pattern={on 10pt off 1pt on 0pt off 0pt} , line width=1.3pt, forget plot]
  table[row sep=crcr]{%
-10	0.858814518385577\\
-9	0.82683407771534\\
-8	0.788869555081048\\
-7	0.744402815659938\\
-6	0.693152327318217\\
-5	0.635203156398987\\
-4	0.571140597515604\\
-3	0.502158246114696\\
-2	0.430100816531021\\
-1	0.357400045874606\\
0	0.286875758506035\\
1	0.221406982956213\\
2	0.163524323967923\\
3	0.115017779564475\\
4	0.0766707640145398\\
5	0.0482045241511244\\
6	0.0284508016426069\\
7	0.0156916877395017\\
8	0.00805215897036482\\
9	0.00382834991314862\\
10	0.00167978287353876\\
};
\addplot [color=mycolor2, dash pattern={on 10pt off 1pt on 1pt off 1pt} , line width=1.3pt, forget plot]
  table[row sep=crcr]{%
-10	0.856241414248259\\
-9	0.823745655484584\\
-8	0.785259090903217\\
-7	0.740351671452615\\
-6	0.688899967660624\\
-5	0.631243666113169\\
-4	0.568336916881081\\
-3	0.501845730256684\\
-2	0.434126230776465\\
-1	0.368023777141906\\
0	0.306477751036544\\
1	0.252002929771819\\
2	0.206210401929661\\
3	0.169559608341723\\
4	0.141446540249038\\
5	0.120566630391402\\
6	0.105363336449841\\
7	0.0943779502329778\\
8	0.0864243041823882\\
9	0.0806184967909921\\
10	0.0763334938063183\\
};
\addplot [color=black, line width=1.3pt, forget plot]
  table[row sep=crcr]{%
-10	0.683627264570501\\
-9	0.621840703992344\\
-8	0.553044402166861\\
-7	0.478598236954184\\
-6	0.400756993117979\\
-5	0.322630307573225\\
-4	0.247916503126111\\
-3	0.180387595725046\\
-2	0.123203849651223\\
-1	0.078252607349542\\
0	0.0457642813489221\\
1	0.0243879229903637\\
2	0.0117130313515176\\
3	0.00501129533981982\\
4	0.00188621656881366\\
5	0.000616162015622784\\
6	0.000172090688546691\\
7	4.04137730478122e-05\\
8	7.83185765773965e-06\\
9	1.22621641029626e-06\\
10	1.51437324407456e-07\\
};
\addplot [color=mycolor1, dash pattern={on 10pt off 1pt on 0pt off 0pt} , line width=1.3pt, forget plot]
  table[row sep=crcr]{%
-10	0.683553739087158\\
-9	0.621714764768621\\
-8	0.552836194244623\\
-7	0.478268387750514\\
-6	0.400260362197292\\
-5	0.321926227696113\\
-4	0.246986200571401\\
-3	0.179254668231864\\
-2	0.121947266944434\\
-1	0.0769990070156746\\
0	0.0446539583304436\\
1	0.0235265667684303\\
2	0.0111360128234294\\
3	0.00468257665300886\\
4	0.00172969378407462\\
5	0.000555167501711806\\
6	0.000153192731320528\\
7	3.59738718316294e-05\\
8	7.11968713440481e-06\\
9	1.17673913269112e-06\\
10	1.61029267142787e-07\\
};
\addplot [color=mycolor2, dash pattern={on 10pt off 1pt on 1pt off 1pt} , line width=1.3pt, forget plot]
  table[row sep=crcr]{%
-10	0.68370220403367\\
-9	0.62219170255007\\
-8	0.553904403445427\\
-7	0.480325564207189\\
-6	0.403856753381882\\
-5	0.327738253430222\\
-4	0.255713007556918\\
-3	0.191422490160309\\
-2	0.137659225609128\\
-1	0.0957403281757092\\
0	0.0652955922753142\\
1	0.0445960281723236\\
2	0.0312553191363294\\
3	0.0229427688110295\\
4	0.0178149518840087\\
5	0.014613639085119\\
6	0.0125603247356363\\
7	0.0111985502293351\\
8	0.0102649501865399\\
9	0.00960582861857218\\
10	0.00912897925648903\\
};
\end{axis}
\end{tikzpicture}%
    \caption{Main results of uplink Poisson cellular scenario for different values of UEs density ($\lambda$). Exact closed form result is compared with Lognormal and Gaussian interference models.}
     \label{fig5}
\end{figure}
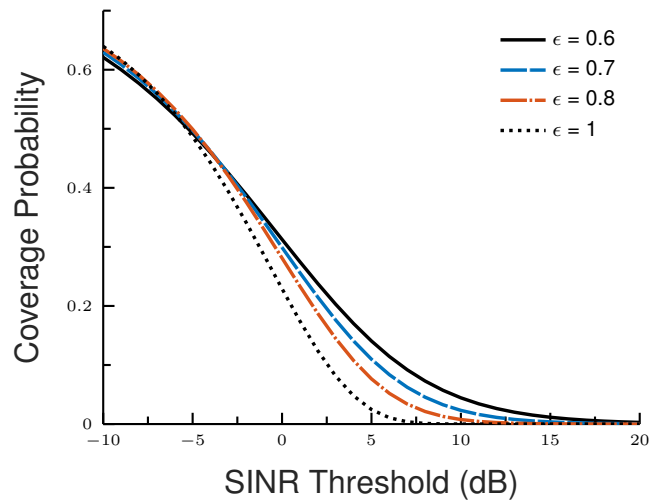
\begin{figure}[H]
\centering
\setlength\fheight{5.5cm}
\setlength\fwidth{7.5cm}
%
%
\definecolor{mycolor1}{rgb}{0.00000,0.44700,0.74100}%
\definecolor{mycolor2}{rgb}{0.85000,0.33000,0.10000}%
\begin{tikzpicture}

\begin{axis}[%
width=0.951\fwidth,
height=\fheight,
at={(0\fwidth,0\fheight)},
scale only axis,
xmin=-10,
xmax=20,
xlabel style={font=\color{white!15!black}},
xlabel={\sffamily{SINR Threshold (dB)}},
ymin=0,
ymax=0.7,
axis x line*=bottom,
axis y line*=left,
axis line style = thick,
every major tick/.append style={thick, black},
every minor tick/.append style={thick, black},
ylabel style={font=\color{white!15!black}},
ylabel={\sffamily{Coverage Probability}},
axis background/.style={fill=white},
legend style={legend cell align=left, align=left, draw=white!15!black,draw=none}
]
\addplot [color=black, line width=1.3pt]
  table[row sep=crcr]{%
-10	0.62102153856515\\
-9	0.600101646213049\\
-8	0.576848787379551\\
-7	0.551171469777184\\
-6	0.523036923617181\\
-5	0.492490152234053\\
-4	0.459672754123239\\
-3	0.424839388209621\\
-2	0.388369248918862\\
-1	0.350769449629629\\
0	0.312666823498389\\
1	0.274784645097254\\
2	0.237901870010685\\
3	0.202795534752255\\
4	0.170172092485963\\
5	0.140599180266193\\
6	0.114452588493754\\
7	0.0918912139523737\\
8	0.0728652922629839\\
9	0.0571533971917951\\
10	0.0444162461644475\\
11	0.0342532967654337\\
12	0.0262511581034165\\
13	0.0200182874805116\\
14	0.015205448719077\\
15	0.0115144717335458\\
16	0.00869891449547228\\
17	0.00655997352275924\\
18	0.00494016918582975\\
19	0.00371645031377204\\
20	0.0027936497248325\\
};
\addlegendentry{\sffamily{$\epsilon\text{ = 0.6}$}}

\addplot [color=mycolor1, dash pattern={on 10pt off 1pt on 0pt off 0pt} , line width=1.3pt]
  table[row sep=crcr]{%
-10	0.629596223952264\\
-9	0.608710376092276\\
-8	0.585195522129134\\
-7	0.558864473801157\\
-6	0.529582024228452\\
-5	0.497294352987728\\
-4	0.462063632236601\\
-3	0.424105347521963\\
-2	0.383823518783488\\
-1	0.341836190078507\\
0	0.29898090411828\\
1	0.256288732350711\\
2	0.214917802079875\\
3	0.176045116765078\\
4	0.140729011809418\\
5	0.109770119951868\\
6	0.0836081756814355\\
7	0.0622867743302093\\
8	0.045496567123478\\
9	0.0326789597307401\\
10	0.0231531079004308\\
11	0.0162287934875738\\
12	0.0112831105941439\\
13	0.00779771441743307\\
14	0.00536572944602808\\
15	0.00368094628436638\\
16	0.00251975365663473\\
17	0.00172230235314823\\
18	0.00117601780379941\\
19	0.00080243758021453\\
20	0.00054726553232677\\
};
\addlegendentry{\sffamily{$\epsilon\text{ = 0.7}$}}

\addplot [color=mycolor2, dash pattern={on 10pt off 1pt on 1pt off 1pt} , line width=1.3pt]
  table[row sep=crcr]{%
-10	0.636005722829001\\
-9	0.614983180086118\\
-8	0.591045720063755\\
-7	0.563912313803955\\
-6	0.533342602624713\\
-5	0.499174557206739\\
-4	0.461374002271281\\
-3	0.420094990400188\\
-2	0.375746099613296\\
-1	0.329051647384091\\
0	0.281089088763805\\
1	0.233276885063032\\
2	0.187285941734193\\
3	0.144859350314453\\
4	0.107554632591901\\
5	0.0764653020887611\\
6	0.0520135935959259\\
7	0.0339026870276055\\
8	0.021258180460732\\
9	0.0129003094682897\\
10	0.00763049002147085\\
11	0.00443048667706137\\
12	0.00254048693394141\\
13	0.00144519293928158\\
14	0.000818146569529895\\
15	0.00046184362867174\\
16	0.000260279818782078\\
17	0.000146546483070005\\
18	8.24664341377448e-05\\
19	4.63924449912521e-05\\
20	2.60941399177203e-05\\
};
\addlegendentry{\sffamily{$\epsilon\text{ = 0.8}$}}

\addplot [color=black, dotted, line width=1.3pt]
  table[row sep=crcr]{%
-10	0.639620464568069\\
-9	0.617001448770834\\
-8	0.590794346829541\\
-7	0.560566378584889\\
-6	0.525925695931761\\
-5	0.486582329737604\\
-4	0.442434949765463\\
-3	0.393684690719612\\
-2	0.340969256603298\\
-1	0.285495686398848\\
0	0.229128068272601\\
1	0.174361859603097\\
2	0.124104255519349\\
3	0.0812077664659419\\
4	0.0478013938828787\\
5	0.0246267858884539\\
6	0.0107272797062901\\
7	0.00378245476951975\\
8	0.00102196615781838\\
9	0.000197457715622611\\
10	2.50095069070255e-05\\
11	1.86105618874029e-06\\
12	7.08836910597425e-08\\
13	1.1616252046312e-09\\
14	6.58258371590971e-12\\
15	9.79578330128813e-15\\
16	2.70743348013695e-18\\
17	8.98602155952031e-23\\
18	2.06847813574845e-28\\
19	1.65448636083122e-35\\
20	1.92640952406659e-44\\
};
\addlegendentry{\sffamily{$\epsilon\text{ = 1}$}}

\end{axis}
\end{tikzpicture}%
    \caption{Effects of the fractional power control on the coverage probability.}
     \label{fig6}
\end{figure}
Fig.~\ref{fig5} presents the variations of the coverage probability for different UE densities. The coverage decreases due to the severity of the interference that increases with interfering UE density. In addition, the Lognormal approximation to model the interference is more accurate than the Gaussian approximation.

In Fig.~\ref{fig6}, we presents the performance of the coverage for different values of fractional power control parameter. Recall that the motivation behind the power control in uplink cellular network is to assist the cell-edge users and improve their coverages. For low SINR threshold $\leq$ -5 dB, the difference in coverage is negligible for all fractional power control $\epsilon$ values. As $\epsilon$ increases, the probability of coverage decreases in particular for $\epsilon = 0.8$ providing much lower coverage between 5 and 10 dB of SINR threshold. Moreover, full pathloss inversion, i.e., unit fractional power control yields a pronounced reduction in coverage accross all the target SINR thresholds. 
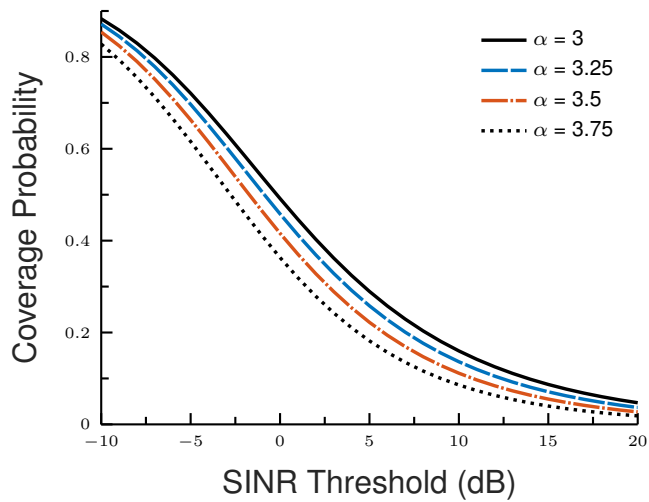
\begin{figure}[H]
\centering
\setlength\fheight{5.5cm}
\setlength\fwidth{7.5cm}
%
%
\definecolor{mycolor1}{rgb}{0.00000,0.44700,0.74100}%
\definecolor{mycolor2}{rgb}{0.85000,0.33000,0.10000}%
\begin{tikzpicture}

\begin{axis}[%
width=0.951\fwidth,
height=\fheight,
at={(0\fwidth,0\fheight)},
scale only axis,
xmin=-10,
xmax=20,
xlabel style={font=\color{white!15!black}},
xlabel={\sffamily{SINR Threshold (dB)}},
ymin=0,
ymax=0.9,
ylabel style={font=\color{white!15!black}},
ylabel={\sffamily{Coverage Probability}},
axis x line*=bottom,
axis y line*=left,
axis line style = thick,
every major tick/.append style={thick, black},
every minor tick/.append style={thick, black},
axis background/.style={fill=white},
legend style={legend cell align=left, align=left, draw=white!15!black,draw=none}
]
\addplot [color=black, line width=1.3pt]
  table[row sep=crcr]{%
-10	0.883026471830464\\
-9	0.85837325261123\\
-8	0.829838414868407\\
-7	0.797326161863694\\
-6	0.760924283450121\\
-5	0.72093666437959\\
-4	0.67789288464641\\
-3	0.632527564373153\\
-2	0.58572830780422\\
-1	0.538459279323506\\
0	0.491674319787035\\
1	0.446236284451567\\
2	0.402856742265757\\
3	0.362063597505393\\
4	0.324196507148439\\
5	0.289424052792742\\
6	0.257773951710129\\
7	0.22916794885559\\
8	0.203455203233383\\
9	0.180440633988856\\
10	0.159906930020996\\
11	0.141630404961527\\
12	0.12539163435639\\
13	0.110982048850926\\
14	0.0982075965288587\\
15	0.0868903962991099\\
16	0.0768690842252265\\
17	0.0679983564272334\\
18	0.0601480537129865\\
19	0.0532020154145833\\
20	0.0470568468296767\\
};
\addlegendentry{\sffamily{$\alpha\text{ = 3}$}}

\addplot [color=mycolor1,dash pattern={on 10pt off 1pt on 0pt off 0pt}, line width=1.3pt]
  table[row sep=crcr]{%
-10	0.871491261628225\\
-9	0.844640337346159\\
-8	0.813677186851935\\
-7	0.778557190956915\\
-6	0.739447545380862\\
-5	0.696757471734021\\
-4	0.651139798838307\\
-3	0.603457015205077\\
-2	0.554712952857565\\
-1	0.505960951579476\\
0	0.45820632280513\\
1	0.412322248805445\\
2	0.368993344558519\\
3	0.328692398657135\\
4	0.291687105510365\\
5	0.258067937676437\\
6	0.227786675497268\\
7	0.200696646377731\\
8	0.176588760753443\\
9	0.155220500708241\\
10	0.136337316321241\\
11	0.119687229156818\\
12	0.105030004463051\\
13	0.092142316082187\\
14	0.0808201397796182\\
15	0.0708793389895074\\
16	0.0621551427885145\\
17	0.0545009972750557\\
18	0.0477871065200812\\
19	0.0418988621124674\\
20	0.0367352808089539\\
};
\addlegendentry{\sffamily{$\alpha\text{ = 3.25}$}}

\addplot [color=mycolor2, dash pattern={on 10pt off 1pt on 1pt off 1pt} , line width=1.3pt]
  table[row sep=crcr]{%
-10	0.854203176583224\\
-9	0.824287050125809\\
-8	0.790029451664851\\
-7	0.75148733753838\\
-6	0.708964727728193\\
-5	0.663033539749868\\
-4	0.61451818548305\\
-3	0.564439660966852\\
-2	0.513925135738707\\
-1	0.464099630796741\\
0	0.415981837090649\\
1	0.370404231340979\\
2	0.327969335344917\\
3	0.289043217737799\\
4	0.253778570302965\\
5	0.222155444015577\\
6	0.194027977408054\\
7	0.1691684858233\\
8	0.14730409970769\\
9	0.12814436369964\\
10	0.111400295304874\\
11	0.0967964125976999\\
12	0.0840774957049127\\
13	0.0730116866123158\\
14	0.0633912070937825\\
15	0.0550316321006446\\
16	0.047770363056311\\
17	0.041464721468955\\
18	0.0359899219667731\\
19	0.0312370894953108\\
20	0.0271113748614835\\
};
\addlegendentry{\sffamily{$\alpha\text{ = 3.5}$}}

\addplot [color=black, dotted, line width=1.3pt]
  table[row sep=crcr]{%
-10	0.828015812354661\\
-9	0.793831773521085\\
-8	0.755140332417614\\
-7	0.712183438003518\\
-6	0.665488125936879\\
-5	0.615864426510954\\
-4	0.564358899136831\\
-3	0.512166048714277\\
-2	0.460512372974697\\
-1	0.410536628036113\\
0	0.363190856045971\\
1	0.319179365941989\\
2	0.278940846602283\\
3	0.242667458953655\\
4	0.210348030188348\\
5	0.181821298535595\\
6	0.156827997528429\\
7	0.135055012935203\\
8	0.116168959998579\\
9	0.0998393775430742\\
10	0.0857532227116682\\
11	0.0736228127368207\\
12	0.063189226389083\\
13	0.0542227889458772\\
14	0.0465218306039119\\
15	0.0399105301552738\\
16	0.0342363654488905\\
17	0.0293674866061928\\
18	0.0251901907359242\\
19	0.0216065895607211\\
20	0.0185325081794785\\
};
\addlegendentry{\sffamily{$\alpha\text{ = 3.75}$}}

\end{axis}
\end{tikzpicture}%
    \caption{Effects of the pathloss exponent on the coverage probability.}
     \label{fig7}
\end{figure}
Fig.~\ref{fig7} illustrates the impacts of the pathloss exponent on the probability of coverage. Basically, the coverage shows significant reduction for higher pathloss exponent around a 0 dB of SINR threshold, in particular between 3 and 3.75 pathloss exponents. Across the higher SINR thresholds, the impacts of the pathloss are minor as the coverage probability converges to the same value regardless of the pathloss exponent values.
\begin{figure}[H]
\centering
\setlength\fheight{5.5cm}
\setlength\fwidth{7.5cm}
%
%
\definecolor{mycolor1}{rgb}{0.00000,0.44700,0.74100}%
\definecolor{mycolor2}{rgb}{0.85000,0.33000,0.10000}%
\begin{tikzpicture}

\begin{axis}[%
width=0.951\fwidth,
height=\fheight,
at={(0\fwidth,0\fheight)},
scale only axis,
xmin=-10,
xmax=80,
xlabel style={font=\color{white!15!black}},
xlabel={\sffamily{Average SNR (dB)}},
ymin=0,
ymax=1,
ylabel style={font=\color{white!15!black}},
ylabel={\sffamily{Rate Coverage}},
axis background/.style={fill=white},
axis x line*=bottom,
axis y line*=left,
axis line style = thick,
every major tick/.append style={thick, black},
every minor tick/.append style={thick, black},
legend style={at={(0.97,0.03)}, anchor=south east, legend cell align=left, align=left, draw=white!15!black,draw=none}
]
\addplot [color=black, line width=1.3pt]
  table[row sep=crcr]{%
-10	8.16473779222671e-05\\
-9	0.000521933964630739\\
-8	0.0024088303786296\\
-7	0.00841844987029239\\
-6	0.0232484052599159\\
-5	0.0526891096042322\\
-4	0.101319312454858\\
-3	0.170188140865785\\
-2	0.256036189639608\\
-1	0.352405104976306\\
0	0.451738648609308\\
1	0.547304845012759\\
2	0.634286501280041\\
3	0.71001285055948\\
4	0.773636179181547\\
5	0.825590227270902\\
6	0.867058336086035\\
7	0.899556611074131\\
8	0.924653713902493\\
9	0.943807774788172\\
10	0.958288334386618\\
11	0.969153013367347\\
12	0.97725553744175\\
13	0.983269141155476\\
14	0.987715421282771\\
15	0.990993040742295\\
16	0.993403515991412\\
17	0.995173033977284\\
18	0.996470196838787\\
19	0.997420062904905\\
20	0.998115037733497\\
21	0.99862319784874\\
22	0.998994581979793\\
23	0.999265906149946\\
24	0.999464074858268\\
25	0.999608782967322\\
26	0.999714436587205\\
27	0.999791567169679\\
28	0.999847870272217\\
29	0.999888967363928\\
30	0.999918963825861\\
31	0.999940857288979\\
32	0.999956836241951\\
33	0.999968498281453\\
34	0.999977009567387\\
35	0.999983221288053\\
36	0.999987754707266\\
37	0.999991063258949\\
38	0.999993477878264\\
39	0.999995240091995\\
40	0.999996526172027\\
41	0.999997464764712\\
42	0.999998149757931\\
43	0.999998649672126\\
44	0.999999014514157\\
45	0.99999928077944\\
46	0.999999475102612\\
47	0.999999616921832\\
48	0.999999720423185\\
49	0.999999795959812\\
50	0.999999851087491\\
};
\addlegendentry{\sffamily{$\text{Clear Air (}\sigma\text{ = 0.43 dB/Km)}$}}

\addplot [color=mycolor1, dash pattern={on 10pt off 1pt on 0pt off 0pt} , line width=1.3pt]
  table[row sep=crcr]{%
-10	3.58814089551629e-11\\
-9	6.97841784358388e-12\\
-8	4.55847470881565e-10\\
-7	1.58835699037851e-08\\
-6	3.21278585402318e-07\\
-5	4.06651795514801e-06\\
-4	3.44182685813044e-05\\
-3	0.000206542962551737\\
-2	0.000925155696418933\\
-1	0.00323488609107692\\
0	0.00917794494191881\\
1	0.0218399988322274\\
2	0.0448317578350542\\
3	0.0812911653171804\\
4	0.132826675425633\\
5	0.198879015204045\\
6	0.276745477277148\\
7	0.362188765889332\\
8	0.450341304970387\\
9	0.536584379732954\\
10	0.617181095331236\\
11	0.68958372081954\\
12	0.752450549325328\\
13	0.805466901475818\\
14	0.849074047525231\\
15	0.884188305383536\\
16	0.911960506489681\\
17	0.933596895613\\
18	0.950242259147943\\
19	0.962915153666264\\
20	0.972481346766556\\
21	0.979652176353478\\
22	0.984997138284973\\
23	0.988963086000905\\
24	0.991895159627065\\
25	0.994056654869212\\
26	0.995646480929459\\
27	0.996813756676332\\
28	0.997669602362904\\
29	0.998296432691353\\
30	0.998755147576706\\
31	0.999090619928848\\
32	0.999335840077591\\
33	0.999515020703224\\
34	0.999645908756777\\
35	0.999741498816779\\
36	0.999811298270321\\
37	0.999862258964519\\
38	0.999899461798117\\
39	0.99992661895252\\
40	0.99994644188149\\
41	0.999960910685089\\
42	0.999971471152718\\
43	0.99997917881771\\
44	0.999984804228784\\
45	0.99998890985518\\
46	0.999991906255411\\
47	0.999994093093308\\
48	0.999995689084928\\
49	0.999996853861229\\
50	0.99999770392764\\
};
\addlegendentry{\sffamily{$\text{Haze (}\sigma\text{ = 4.2 dB/Km)}$}}

\addplot [color=mycolor2, dash pattern={on 10pt off 1pt on 1pt off 1pt} , line width=1.3pt]
  table[row sep=crcr]{%
-10	0\\
-9	0\\
-8	0\\
-7	0\\
-6	2.98383540098257e-12\\
-5	1.76740400092967e-10\\
-4	5.733146068998e-09\\
-3	1.10990167723202e-07\\
-2	1.3812042052308e-06\\
-1	1.17850875495096e-05\\
0	7.28964769135532e-05\\
1	0.000342868555816334\\
2	0.00127721659156821\\
3	0.00390011603999563\\
4	0.0100505393955302\\
5	0.0223996737969384\\
6	0.0440781892683231\\
7	0.0779405663228282\\
8	0.125711866763774\\
9	0.187356599550583\\
10	0.260918440120438\\
11	0.342870804300229\\
12	0.428819274469881\\
13	0.514301922655371\\
14	0.595458623787747\\
15	0.669439685555981\\
16	0.73453477715145\\
17	0.790081317523833\\
18	0.836243509395833\\
19	0.873747982876486\\
20	0.903637189723803\\
21	0.927073019919914\\
22	0.945199946891192\\
23	0.959062491455938\\
24	0.969565124264083\\
25	0.97746148586336\\
26	0.98336151112323\\
27	0.98774784988896\\
28	0.990995777609747\\
29	0.993393083500813\\
30	0.995158081182946\\
31	0.996454970799301\\
32	0.997406424981794\\
33	0.998103609892188\\
34	0.998613998669876\\
35	0.998987368941189\\
36	0.999260352188909\\
37	0.999459853600576\\
38	0.999605605111963\\
39	0.999712061262431\\
40	0.999789801309394\\
41	0.999846562951642\\
42	0.999888002627858\\
43	0.999918253686357\\
44	0.999940335587219\\
45	0.999956453571809\\
46	0.999968217938588\\
47	0.999976804392251\\
48	0.999983071245741\\
49	0.999987645054062\\
50	0.999990983165384\\
};
\addlegendentry{\sffamily{$\text{Moderate Rain (}\sigma\text{ = 5.8 dB/Km)}$}}

\addplot [color=black, dotted, line width=1.3pt]
  table[row sep=crcr]{%
-10	0\\
-9	0\\
-8	0\\
-7	0\\
-6	0\\
-5	0\\
-4	0\\
-3	0\\
-2	3.06560332674621e-11\\
-1	9.86707049399627e-10\\
0	1.96029488197524e-08\\
1	2.57818878135474e-07\\
2	2.38196570734317e-06\\
3	1.62529618944474e-05\\
4	8.54299557708771e-05\\
5	0.000358397595675464\\
6	0.00123643932058637\\
7	0.00359763027189519\\
8	0.00902098052386968\\
9	0.0198573804639458\\
10	0.0389931079949994\\
11	0.0692702145384015\\
12	0.112714063765626\\
13	0.169851295341372\\
14	0.239388630360357\\
15	0.318371571715657\\
16	0.402746268434124\\
17	0.488112949180259\\
18	0.570435454585886\\
19	0.646540116074407\\
20	0.714343095887168\\
21	0.772836197017041\\
22	0.821910846606047\\
23	0.862108512489198\\
24	0.894367855303537\\
25	0.919811282978019\\
26	0.939588197353502\\
27	0.954774555366132\\
28	0.966318969021671\\
29	0.975022514167371\\
30	0.981540186229575\\
31	0.986394428213443\\
32	0.989993992932611\\
33	0.992653881465851\\
34	0.994613973689203\\
35	0.996055239265104\\
36	0.997113202466908\\
37	0.997888769208552\\
38	0.998456730664229\\
39	0.998872326395636\\
40	0.999176243524814\\
41	0.999398386682057\\
42	0.999560699290455\\
43	0.999679262549378\\
44	0.999765850019552\\
45	0.999829075031525\\
46	0.99987523532084\\
47	0.999908933520955\\
48	0.999933532285133\\
49	0.999951487713569\\
50	0.999964593403086\\
};
\addlegendentry{\sffamily{$\text{Heavy Rain (}\sigma\text{ = 9.2 dB/Km)}$}}

\end{axis}
\end{tikzpicture}%
    \caption{Impacts of the atmospheric pathloss on the rate coverage.}
     \label{fig8}
\end{figure}
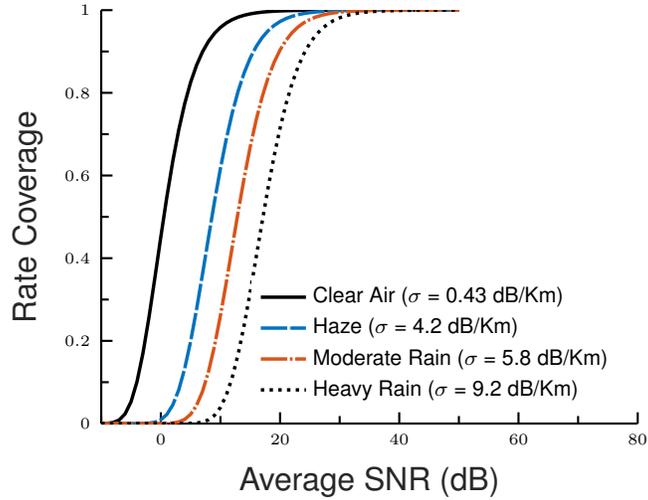

Fig.~\ref{fig8} provides the dependence of the rate coverage on the atmospheric pathloss at the second hop. The best performance are achieved for clear air weather as the attenuation factor is small. Across severe values of the pathloss, the coverage gradually degrades in particular for rainy weather. In fact, the scattering process becomes significant with the rainfall intensity when the signal hits the rain droplets. Consequently, the signal experiences scattering in different directions which introduces a significant loss in signal power leading to diminishing the rate coverage.
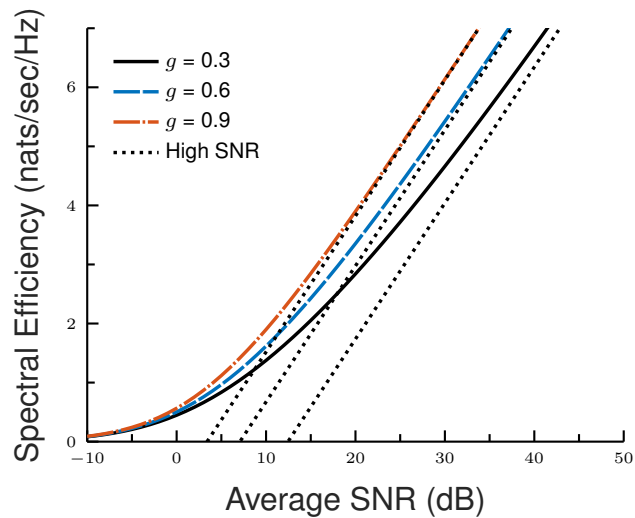
\begin{figure}[H]
\centering
\setlength\fheight{5.5cm}
\setlength\fwidth{7.5cm}
%
%
\definecolor{mycolor1}{rgb}{0.00000,0.44700,0.74100}%
\definecolor{mycolor2}{rgb}{0.85000,0.33000,0.10000}%
\begin{tikzpicture}

\begin{axis}[%
width=0.951\fwidth,
height=\fheight,
at={(0\fwidth,0\fheight)},
scale only axis,
xmin=-10,
xmax=50,
xlabel style={font=\color{white!15!black}},
xlabel={\sffamily{Average SNR (dB)}},
ymin=0,
ymax=7,
axis x line*=bottom,
axis y line*=left,
axis line style = thick,
every major tick/.append style={thick, black},
every minor tick/.append style={thick, black},
ylabel style={font=\color{white!15!black}},
ylabel={\sffamily{Spectral Efficiency (nats/sec/Hz)}},
axis background/.style={fill=white},
legend style={at={(0.03,0.97)}, anchor=north west, legend cell align=left, align=left, draw=white!15!black,draw=none}
]
\addplot [color=black, line width=1.3pt]
  table[row sep=crcr]{%
-10	0.0826595816164354\\
-9	0.100391575108637\\
-8	0.12131192937255\\
-7	0.145806352393909\\
-6	0.174261635521426\\
-5	0.207056479862932\\
-4	0.244552279798246\\
-3	0.287084427066278\\
-2	0.334954644128645\\
-1	0.388424749211948\\
0	0.447712115908212\\
1	0.512986938845441\\
2	0.584371273828429\\
3	0.661939701529169\\
4	0.745721377541763\\
5	0.835703181286977\\
6	0.931833659433561\\
7	1.03402747020407\\
8	1.1421700654302\\
9	1.25612238963365\\
10	1.3757254227867\\
11	1.50080444033618\\
12	1.63117290684003\\
13	1.76663595602463\\
14	1.9069934393601\\
15	2.05204254743387\\
16	2.20158002413638\\
17	2.35540400392615\\
18	2.51331550828947\\
19	2.67511963999257\\
20	2.8406265137704\\
21	3.0096519604801\\
22	3.18201803908751\\
23	3.35755338762082\\
24	3.53609344075822\\
25	3.71748053825219\\
26	3.90156394508691\\
27	4.08819980120875\\
28	4.27725101591141\\
29	4.46858711951232\\
30	4.66208408282231\\
31	4.8576241130684\\
32	5.05509543335559\\
33	5.2543920514204\\
34	5.45541352230736\\
35	5.65806470866325\\
36	5.86225554156519\\
37	6.06790078415649\\
38	6.27491979983548\\
39	6.48323632631122\\
40	6.69277825648856\\
41	6.90347742686152\\
42	7.11526941386533\\
43	7.32809333845581\\
44	7.54189167903996\\
45	7.75661009276862\\
46	7.97219724511402\\
47	8.18860464758765\\
48	8.40578650340322\\
49	8.6236995608524\\
50	8.84230297413471\\
};
\addlegendentry{\sffamily{$g$ = 0.3}}

\addplot [color=mycolor1, dash pattern={on 10pt off 1pt on 0pt off 0pt} , line width=1.3pt]
  table[row sep=crcr]{%
-10	0.0863251734117366\\
-9	0.105518270774442\\
-8	0.128381777003092\\
-7	0.155416214570267\\
-6	0.187134589504663\\
-5	0.224049638493656\\
-4	0.266659947403214\\
-3	0.315435804550581\\
-2	0.370805707723238\\
-1	0.433144393375439\\
0	0.502763109608558\\
1	0.579902637048175\\
2	0.664729308357563\\
3	0.757334024181784\\
4	0.857734041819547\\
5	0.965877144127324\\
6	1.08164768994485\\
7	1.20487400310014\\
8	1.33533656634798\\
9	1.47277653637878\\
10	1.61690417179896\\
11	1.76740685417649\\
12	1.92395647176793\\
13	2.08621601840266\\
14	2.25384533128297\\
15	2.42650594889337\\
16	2.60386511353421\\
17	2.78559897328014\\
18	2.9713950572557\\
19	3.16095410816696\\
20	3.3539913591721\\
21	3.55023734035651\\
22	3.74943829494174\\
23	3.95135627821655\\
24	4.15576900403267\\
25	4.36246949528121\\
26	4.5712655865537\\
27	4.78197931950795\\
28	4.99444626448451\\
29	5.20851479573344\\
30	5.42404534222917\\
31	5.6409096314411\\
32	5.85898993953446\\
33	6.07817835822728\\
34	6.29837608585081\\
35	6.51949274797621\\
36	6.74144575120833\\
37	6.9641596723438\\
38	7.18756568398517\\
39	7.41160101684695\\
40	7.63620845833575\\
41	7.86133588650096\\
42	8.08693583809957\\
43	8.31296510927447\\
44	8.53938438718592\\
45	8.76615791084392\\
46	8.9932531593488\\
47	9.22064056574695\\
48	9.44829325473794\\
49	9.67618680252066\\
50	9.90429901713274\\
};
\addlegendentry{\sffamily{$g$ = 0.6}}

\addplot [color=mycolor2, dash pattern={on 10pt off 1pt on 1pt off 1pt} , line width=1.3pt]
  table[row sep=crcr]{%
-10	0.0898090519262892\\
-9	0.110487177036774\\
-8	0.135371727780164\\
-7	0.16510846727108\\
-6	0.2003748967381\\
-5	0.241863007868342\\
-4	0.290258605817868\\
-3	0.346218351245304\\
-2	0.410346045354308\\
-1	0.483169890098788\\
0	0.565122452025801\\
1	0.656524837370327\\
2	0.757576182809909\\
3	0.868349046794509\\
4	0.988790732057645\\
5	1.11873005995764\\
6	1.25788871377847\\
7	1.40589600735755\\
8	1.56230582624378\\
9	1.72661451649732\\
10	1.89827863093772\\
11	2.07673164665586\\
12	2.26139900377925\\
13	2.45171105229796\\
14	2.64711370811341\\
15	2.84707679726493\\
16	3.05110020246986\\
17	3.25871801887633\\
18	3.46950098065685\\
19	3.68305744344414\\
20	3.89903320708715\\
21	4.11711044596853\\
22	4.33700598644676\\
23	4.55846913793267\\
24	4.78127924949264\\
25	5.00524313034871\\
26	5.23019244190732\\
27	5.45598114191819\\
28	5.68248303840561\\
29	5.90958949210034\\
30	6.13720729095925\\
31	6.36525670858254\\
32	6.59366974945473\\
33	6.82238857746922\\
34	7.05136411969673\\
35	7.28055483442028\\
36	7.50992563073212\\
37	7.73944692617489\\
38	7.96909382876642\\
39	8.19884543008027\\
40	8.42868419670651\\
41	8.6585954482718\\
42	8.88856691116414\\
43	9.11858833811945\\
44	9.34865118483773\\
45	9.57874833577294\\
46	9.80887387216162\\
47	10.0390228762086\\
48	10.2691912661263\\
49	10.4993756574278\\
50	10.7295732464979\\
};
\addlegendentry{\sffamily{$g$ = 0.9}}

\addplot [color=black, dotted, line width=1.3pt]
  table[row sep=crcr]{%
-10	-5.16992335855069\\
-9	-4.93966484925129\\
-8	-4.70940633995188\\
-7	-4.47914783065248\\
-6	-4.24888932135307\\
-5	-4.01863081205367\\
-4	-3.78837230275426\\
-3	-3.55811379345486\\
-2	-3.32785528415545\\
-1	-3.09759677485605\\
0	-2.86733826555665\\
1	-2.63707975625724\\
2	-2.40682124695784\\
3	-2.17656273765843\\
4	-1.94630422835903\\
5	-1.71604571905962\\
6	-1.48578720976022\\
7	-1.25552870046081\\
8	-1.02527019116141\\
9	-0.795011681862003\\
10	-0.564753172562599\\
11	-0.334494663263194\\
12	-0.104236153963789\\
13	0.126022355335615\\
14	0.35628086463502\\
15	0.586539373934424\\
16	0.81679788323383\\
17	1.04705639253323\\
18	1.27731490183264\\
19	1.50757341113204\\
20	1.73783192043145\\
21	1.96809042973085\\
22	2.19834893903026\\
23	2.42860744832966\\
24	2.65886595762907\\
25	2.88912446692847\\
26	3.11938297622788\\
27	3.34964148552728\\
28	3.57989999482669\\
29	3.81015850412609\\
30	4.04041701342549\\
31	4.2706755227249\\
32	4.5009340320243\\
33	4.73119254132371\\
34	4.96145105062311\\
35	5.19170955992252\\
36	5.42196806922192\\
37	5.65222657852133\\
38	5.88248508782073\\
39	6.11274359712014\\
40	6.34300210641954\\
41	6.57326061571895\\
42	6.80351912501835\\
43	7.03377763431775\\
44	7.26403614361716\\
45	7.49429465291656\\
46	7.72455316221597\\
47	7.95481167151538\\
48	8.18507018081478\\
49	8.41532869011418\\
50	8.64558719941359\\
};
\addlegendentry{\sffamily{High SNR}}
\addplot [color=black, dotted, line width=1.3pt, forget plot]
  table[row sep=crcr]{%
-10	-3.93224345710698\\
-9	-3.70198494780757\\
-8	-3.47172643850817\\
-7	-3.24146792920876\\
-6	-3.01120941990936\\
-5	-2.78095091060995\\
-4	-2.55069240131055\\
-3	-2.32043389201115\\
-2	-2.09017538271174\\
-1	-1.85991687341234\\
0	-1.62965836411293\\
1	-1.39939985481353\\
2	-1.16914134551412\\
3	-0.938882836214717\\
4	-0.708624326915313\\
5	-0.478365817615908\\
6	-0.248107308316503\\
7	-0.0178487990170991\\
8	0.212409710282306\\
9	0.442668219581711\\
10	0.672926728881115\\
11	0.90318523818052\\
12	1.13344374747992\\
13	1.36370225677933\\
14	1.59396076607873\\
15	1.82421927537814\\
16	2.05447778467754\\
17	2.28473629397695\\
18	2.51499480327635\\
19	2.74525331257576\\
20	2.97551182187516\\
21	3.20577033117457\\
22	3.43602884047397\\
23	3.66628734977337\\
24	3.89654585907278\\
25	4.12680436837219\\
26	4.35706287767159\\
27	4.58732138697099\\
28	4.8175798962704\\
29	5.0478384055698\\
30	5.27809691486921\\
31	5.50835542416861\\
32	5.73861393346802\\
33	5.96887244276742\\
34	6.19913095206683\\
35	6.42938946136623\\
36	6.65964797066563\\
37	6.88990647996504\\
38	7.12016498926445\\
39	7.35042349856385\\
40	7.58068200786326\\
41	7.81094051716266\\
42	8.04119902646207\\
43	8.27145753576147\\
44	8.50171604506087\\
45	8.73197455436028\\
46	8.96223306365968\\
47	9.19249157295909\\
48	9.42275008225849\\
49	9.6530085915579\\
50	9.8832671008573\\
};
\addplot [color=black, dotted, line width=1.3pt, forget plot]
  table[row sep=crcr]{%
-10	-3.08621409042485\\
-9	-2.85595558112544\\
-8	-2.62569707182604\\
-7	-2.39543856252663\\
-6	-2.16518005322723\\
-5	-1.93492154392782\\
-4	-1.70466303462842\\
-3	-1.47440452532901\\
-2	-1.24414601602961\\
-1	-1.0138875067302\\
0	-0.7836289974308\\
1	-0.553370488131395\\
2	-0.32311197883199\\
3	-0.0928534695325857\\
4	0.137405039766819\\
5	0.367663549066224\\
6	0.597922058365628\\
7	0.828180567665033\\
8	1.05843907696444\\
9	1.28869758626384\\
10	1.51895609556325\\
11	1.74921460486265\\
12	1.97947311416206\\
13	2.20973162346146\\
14	2.43999013276087\\
15	2.67024864206027\\
16	2.90050715135968\\
17	3.13076566065908\\
18	3.36102416995848\\
19	3.59128267925789\\
20	3.82154118855729\\
21	4.0517996978567\\
22	4.2820582071561\\
23	4.51231671645551\\
24	4.74257522575491\\
25	4.97283373505432\\
26	5.20309224435372\\
27	5.43335075365313\\
28	5.66360926295253\\
29	5.89386777225194\\
30	6.12412628155134\\
31	6.35438479085075\\
32	6.58464330015015\\
33	6.81490180944955\\
34	7.04516031874896\\
35	7.27541882804836\\
36	7.50567733734777\\
37	7.73593584664717\\
38	7.96619435594658\\
39	8.19645286524598\\
40	8.42671137454539\\
41	8.65696988384479\\
42	8.88722839314419\\
43	9.1174869024436\\
44	9.34774541174301\\
45	9.57800392104241\\
46	9.80826243034181\\
47	10.0385209396412\\
48	10.2687794489406\\
49	10.49903795824\\
50	10.7292964675394\\
};
\end{axis}
\end{tikzpicture}%
    \caption{Impacts of the pointing errors on the ergodic achievable rate.}
     \label{fig9}
\end{figure}
Fig.~\ref{fig9} shows the variations of the ergodic rate on the pointing errors. Recall that the severity of the pointing errors is inversely proportional to the pointing errors coefficient $g$ considered for the mathematical system model in this work. The figure shows that the performance gets much worse for severe pointing errors and vice versa. In fact, the pointing errors or the misalignment between the transmitter and the front photodector is explained by the laser beam perturbations or fluctuations caused by seismic activities, building swaying or potential blockage by birds, etc. Due to the signal vibrations, the front photodector captures only a fraction of the intended received power yielding a significant losses in the ergodic rate.
\begin{figure}[H]
\centering
\setlength\fheight{5.5cm}
\setlength\fwidth{7.5cm}
\input{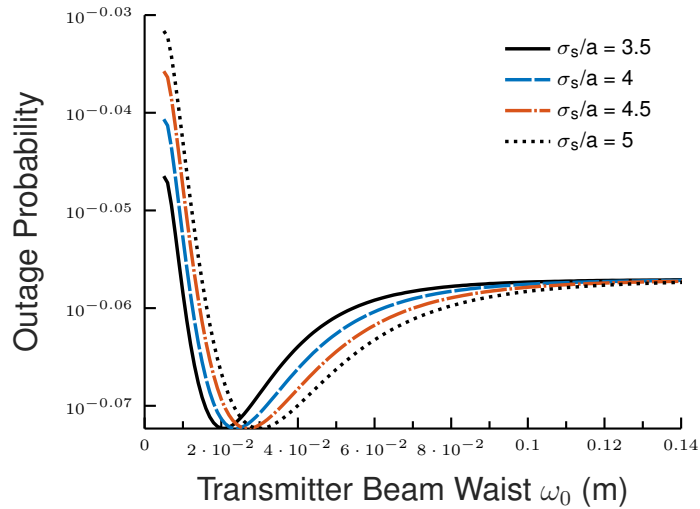}
    \caption{Effects of the transmitter beam waist on the outage probability. The performance is evaluated for different values of the ratio between the Jitter variance and the receiver apertures.}
     \label{fig10}
\end{figure}
Fig.~\ref{fig10} illustrates the variations of the outage probability on the beam waist the BS, for different values of the normalized Jitter standard deviation. We observe that the probability of outage varies for several orders of amplitude across all the values of the beam waist. Specifically, the outage is minimized for a unique optimal value $\omega_{\texttt{0,opt}}$ of the beam waist for a given normalized Jitter variance. However, the mismatch between the beam waist and the optimal value introduces a severe degradation in outage performance. Moreover, the optimal value $\omega_{\texttt{0,opt}}$ shifted to higher values with increasing the normalized Jitter variance. The values of the optimal values of the beam waists at the BS and the data center are summarized by the following TABLE \ref{beamwaist}.
\begin{table}[H]
\renewcommand{\arraystretch}{.8}
\caption{Optimum values of the laser beam waists at the BS and the data center.}
\label{beamwaist}
\centering
\begin{tabular}{|c|c|c|}
\hline\hline
\bfseries $\sigma_s/a$ & \bfseries $\omega_{\texttt{0,opt}}$ [cm] &  \bfseries $\omega_{\texttt{L,opt}}/a$\\
\hline
3.5 & 2.1  & 4.2340\\
\hline
4 & 2.4  & 4.8340\\
\hline
4.5 & 2.7  & 5.4360\\
\hline
5 & 3  & 6.0380 \\
\hline\hline
\end{tabular}
\end{table}
\begin{figure}[H]
\centering
\setlength\fheight{5.5cm}
\setlength\fwidth{7.5cm}
\input{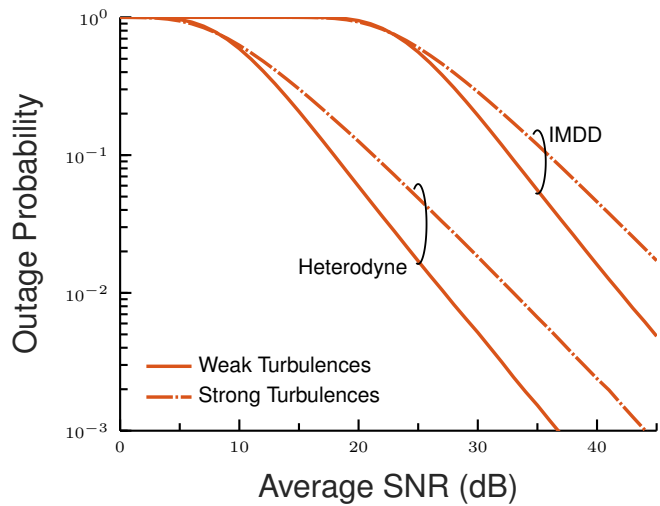}
    \caption{Joint effects of the receiver optical detection and the atmospheric turbulences on the outage probability.}
     \label{fig11}
\end{figure}
Fig.~ \ref{fig11} illustrates the impacts of the demodulation scheme and the atmospheric turbulences on the outage probability. As expected, the system works better under low turbulences and vice versa. In addition, the performance further improves under heterodyne or coherent demodulation compared to IM/DD detection. Furthermore, the impacts of the turbulences-induced fading and the demodulation scheme become more pronounced at high SNR. This can be interpreted by referring to the diversity gain which improves under low turbulences and heterodyne detection as illustrated by (\ref{diversity}).

\begin{figure}[H]
\centering
\setlength\fheight{5.5cm}
\setlength\fwidth{7.5cm}
\input{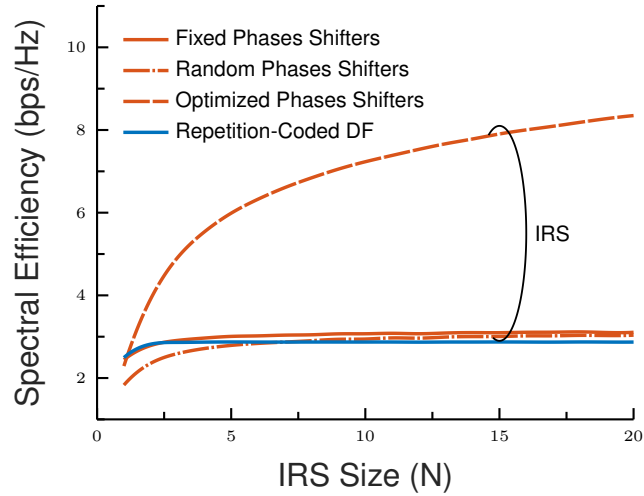}
    \caption{Impacts of the number of reflecting elements on the spectral efficiency for different scenarios.}
     \label{fig12}
\end{figure}

Fig.~\ref{fig12} illustrates the variations of the spectral efficiency with respect to the number of reflecting elements. For random and fixed phase shifters, the spectral efficiency is roughly constant and the system does not exhibit any gain by increasing the number of reflecting elements. In addition, the DF performances are comparable with the last two settings even by increasing the size of IRS. In this case, it is better to operate with DF relaying as it is less costly compared to the IRS. On the other side, the system achieves considerable gain with optimal design of phases shifters compared to other settings. Given that the spectral efficiency is logarithmically increasing with the size of IRS, the effective cost becomes higher with increasing the panel size, and hence it is important to establish a trade-off between the spectral efficiency gain and the effective cost to come up with a practical design. Note that this idea can be considered for the next future work. 
\begin{figure}[H]
\centering
\setlength\fheight{5.5cm}
\setlength\fwidth{7.5cm}
\input{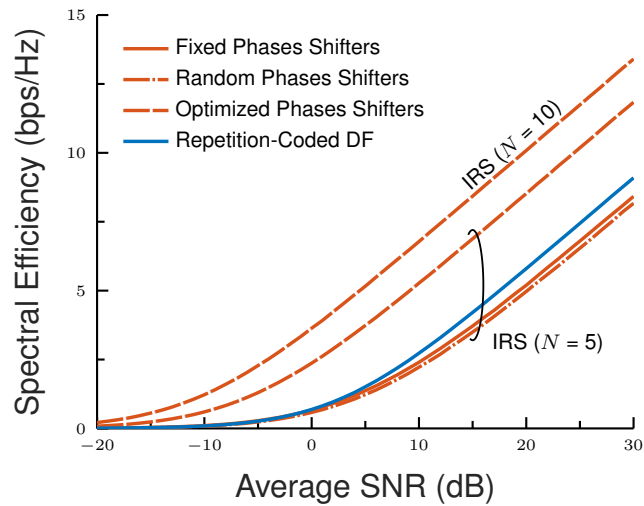}
    \caption{Comparison between repetition-coded DF and IRS setting. For the latter, we consider 3 different designs for the reflecting elements.}
     \label{fig13}
\end{figure}
Fig.~\ref{fig13} presents a comparison between the repetition-coded DF and different configurations of IRS phases shifters. We observe that the optimal design of phase shifters outperforms the other settings in terms of spectral efficiency. In addition, the spectral efficiency for the optimum designs gets much better for high value of reflecting elements ($N = 10$) which is confirmed by the results illustrated by Fig.~\ref{fig12}. Note that by increasing the IRS size, the system does not achieve a multiplexing gain (pre-log factor) rather than there is an SNR offset (power offset) between the two settings for $N = 5$ and $N = 10$ for optimal design.

\section{Conclusion}
In this work, we evaluated the performance of an uplink cellular network with FSO backhauling. We refer to the stochastic geometry framework to model and derive the analytical expressions in terms of SINR statistics, coverage/outage probability and achievable rate of the uplink network wherein FPC is employed to support the cell-edge users and save the overall power consumed by UEs. Independently, we derived the performance expressions of the FSO backhauling where the channel model involves generalized probabilistic models for the atmospheric turbulences and the pointing errors which are distributed following M\'alaga and Beckmann, respectively. Taking advantages of the analysis of the two parts of the network, we evaluated the benchmarking performance of the repetition-coded DF relaying in terms of coverage/outage and ergodic rate. In the same context, we proposed an optimal design of the phase shifters of the IRS and we established a fair comparison between the benchmarking DF relaying and IRS-supported transmission in terms of spectral efficiency. We showed that IRS can beat relaying even for small number of reflecting elements but with robust design of phase shifters in a way to place nulls on the interfering directions while maximizing the received power in LOS direction. Reflectarrays generally have small size compared to conventional relaying, even in case for large number of reflecting elements, as each discrete element has a size in order of sub-wavelength. Thereby, IRS can be a practical solution to be deployed indoor (homes or tunnels) not only because of the compact size of even hundreds of elements but also the cost is too high to be installed everywhere in outdoor. A major concern to reduce the cost is to evaluate the minimum required number of reflecting elements for IRS to beat relaying and maintain affordable costs which was also the objective of the proposed phase shifters design.
\appendices
\section{Proof of the Laplace Transform of The interference}
We start by defining the Laplace Transform of the interference as follows
\begin{equation}\label{LT}
\mathscr{L}_{\mathcal{I}_z}(s) = \mathbb{E}_{I_z}\left[\exp\left(-\sum_{z\in \mathcal{Z}}sr_z^{\alpha\epsilon}h_zd_z^{-\alpha}  \right)  \right].    
\end{equation}
The interference $I_z$ depends on the three random variables $r_z,~g_z,$ and $d_z$. After some mathematical manipulations, (\ref{LT}) can be reformulated sa follows
\begin{equation}\label{LT1}
\mathscr{L}_{\mathcal{I}_z}(s) = \mathbb{E}_{r_z,h_z,d_z}\left[\prod_{z\in \mathcal{Z}}\exp\left( -sr_z^{\alpha\epsilon}h_zd_z^{-\alpha} \right)   \right].    
\end{equation}
Since the channel gains $h_z$ are independent, (\ref{LT1}) is given by
\begin{equation}\label{LT2}
 \mathscr{L}_{\mathcal{I}_z}(s) =  \mathbb{E}_{r_z,d_z}\left[\prod_{z\in \mathcal{Z}}\mathbb{E}_{h_z}\left[\exp\left(-sr_z^{\alpha\epsilon}h_zd_z^{-\alpha} \right)   \right]  \right].
\end{equation}
Given that $h_z$ are exponentially distributed and $r_z$ are independent, (\ref{LT2}) can be expanded as
\begin{equation}\label{LT3}
  \mathscr{L}_{\mathcal{I}_z}(s) = \mathbb{E}_{d_z}\left[\prod_{z\in \mathcal{Z}}\mathbb{E}_{r_z}\left[\frac{\mu}{\mu + sr_z^{\alpha\epsilon}d_z^{-\alpha}} \right] \right]. 
\end{equation}
Referring to the Probability Generating Functional (PGFL) of a PPP \cite{pgfl}, it follows that (\ref{LT3}) can be derived as
\begin{equation}
\mathscr{L}_{\mathcal{I}_z}(s) = \exp\left(-2\pi\lambda \int\limits_r^{\infty}\left(1-\mathbb{E}_{r_z}\left[\frac{\mu}{\mu + sr_z^{\alpha\epsilon}x^{-\alpha}}  \right]   \right)x\text{d}x \right).
\end{equation}

\section{Proof of the Uplink Achievable Rate}
The ergodic rate is defined as follows
\begin{equation}
\begin{split}
\mathcal{I}({\scriptsize{\textsf{SINR}}}) &= \mathbb{E}\left[\log(1+{\scriptsize{\textsf{SINR}}})  \right] \\ &
= \int\limits_0^{+\infty}\int\limits_0^{+\infty} \mathbb{P}\left[\log(1+{\scriptsize{\textsf{SINR}}})>t \right]\text{d}tf_{R}(r)\text{d}r\\ &
= \int\limits_0^{+\infty} f_R(r)\int\limits_0^{+\infty}\mathbb{P}\left[\log\left(1+\frac{hr^{\alpha(\epsilon-1)} }{\sigma^2 + \mathcal{I}_z}  \right)> t \right]\text{d}t\text{d}r \\ &
= \int\limits_0^{+\infty} f_R(r)\int\limits_0^{+\infty} \mathbb{P}\left [h>\frac{(e^t-1)(\sigma^2 + \mathcal{I}_z)}{r^{\alpha(\epsilon-1)}}  \right]\text{d}t\text{d}r\\ &
= \int\limits_0^{+\infty} f_R(r)\int\limits_0^{+\infty} e^{-s\sigma^2}\mathscr{L}_{\mathcal{I}_z}\left(\mu\frac{e^t-1}{r^{\alpha(\epsilon-1)}}  \right)  \text{d}t\text{d}r
\end{split}
\end{equation}
where $f_R(r) = 2\pi\lambda re^{-\pi \lambda r^2},~ r \geq 0$, is the PDF of the nearest BS to the UE which is Rayleigh distributed.

\bibliographystyle{IEEEtran}
\bibliography{main}

\begin{thebibliography}{10}
\providecommand{\url}[1]{#1}
\csname url@samestyle\endcsname
\providecommand{\newblock}{\relax}
\providecommand{\bibinfo}[2]{#2}
\providecommand{\BIBentrySTDinterwordspacing}{\spaceskip=0pt\relax}
\providecommand{\BIBentryALTinterwordstretchfactor}{4}
\providecommand{\BIBentryALTinterwordspacing}{\spaceskip=\fontdimen2\font plus
\BIBentryALTinterwordstretchfactor\fontdimen3\font minus
  \fontdimen4\font\relax}
\providecommand{\BIBforeignlanguage}[2]{{%
\expandafter\ifx\csname l@#1\endcsname\relax
\typeout{** WARNING: IEEEtran.bst: No hyphenation pattern has been}%
\typeout{** loaded for the language `#1'. Using the pattern for}%
\typeout{** the default language instead.}%
\else
\language=\csname l@#1\endcsname
\fi
#2}}
\providecommand{\BIBdecl}{\relax}
\BIBdecl

\bibitem{v1}
J.~{Choi}, V.~{Va}, N.~{Gonzalez-Prelcic}, R.~{Daniels}, C.~R. {Bhat}, and
  R.~W. {Heath}, ``Millimeter-wave vehicular communication to support massive
  automotive sensing,'' \emph{IEEE Communications Magazine}, vol.~54, no.~12,
  pp. 160--167, December 2016.

\bibitem{v2}
P.~{Belanovic}, D.~{Valerio}, A.~{Paier}, T.~{Zemen}, F.~{Ricciato}, and C.~F.
  {Mecklenbrauker}, ``On wireless links for vehicle-to-infrastructure
  communications,'' \emph{IEEE Transactions on Vehicular Technology}, vol.~59,
  no.~1, pp. 269--282, Jan 2010.

\bibitem{cu}
O.~{Tipmongkolsilp}, S.~{Zaghloul}, and A.~{Jukan}, ``The evolution of cellular
  backhaul technologies: Current issues and future trends,'' \emph{IEEE
  Communications Surveys Tutorials}, vol.~13, no.~1, pp. 97--113, First 2011.

\bibitem{surv}
M.~A. {Khalighi} and M.~{Uysal}, ``Survey on free space optical communication:
  A communication theory perspective,'' \emph{IEEE Communications Surveys
  Tutorials}, vol.~16, no.~4, pp. 2231--2258, Fourthquarter 2014.

\bibitem{trinh}
P.~V. {Trinh}, T.~{Cong Thang}, and A.~T. {Pham}, ``Mixed mmwave rf/fso
  relaying systems over generalized fading channels with pointing errors,''
  \emph{IEEE Photonics Journal}, vol.~9, no.~1, pp. 1--14, Feb 2017.

\bibitem{bend}
X.~{Jiao}, H.~{Zhang}, H.~{Li}, X.~{Zhang}, L.~{Xi}, and Z.~{Zhang},
  ``Macro-bending losses of circular photonic crystal fiber supporting 14 oam
  modes,'' in \emph{2018 Asia Communications and Photonics Conference (ACP)},
  Oct 2018, pp. 1--3.

\bibitem{of4}
S.~K. {Korotky}, ``Price-points for components of multi-core fiber
  communication systems in backbone optical networks,'' \emph{IEEE/OSA Journal
  of Optical Communications and Networking}, vol.~4, no.~5, pp. 426--435, May
  2012.

\bibitem{uysal}
{Murat Uysal, Carlo Capsoni, Zabih Ghassemlooy, Anthony Boucouvalas, Eszter
  Udvary}, \emph{Optical Wireless Communications: An Emerging Technology},
  1st~ed., ser. Signals and Communication Technology.\hskip 1em plus 0.5em
  minus 0.4em\relax Springer International Publishing, 2016.

\bibitem{nasab}
E.~Soleimani-Nasab and M.~Uysal, ``Generalized performance analysis of mixed
  {{RF/FSO}} cooperative systems,'' \emph{IEEE Transactions on Wireless
  Communications}, vol.~15, no.~1, pp. 714--727, Jan 2016.

\bibitem{zedini}
E.~{Zedini}, H.~{Soury}, and M.~{Alouini}, ``On the performance analysis of
  dual-hop mixed fso/rf systems,'' \emph{IEEE Transactions on Wireless
  Communications}, vol.~15, no.~5, pp. 3679--3689, May 2016.

\bibitem{d1}
M.~{Safari} and M.~{Uysal}, ``Relay-assisted free-space optical
  communication,'' \emph{IEEE Transactions on Wireless Communications}, vol.~7,
  no.~12, pp. 5441--5449, December 2008.

\bibitem{d2}
Y.~{Li}, M.~{Pióro}, and V.~{Angelakisi}, ``Design of cellular backhaul
  topology using the fso technology,'' in \emph{2013 2nd International Workshop
  on Optical Wireless Communications (IWOW)}, Oct 2013, pp. 6--10.

\bibitem{d3}
{Jifang Zhuang}, M.~J. {Casey}, S.~D. {Milner}, S.~A. {Gabriel}, and G.~B.
  {Baecher}, ``Multi-objective optimization techniques in topology control of
  free space optical networks,'' in \emph{IEEE MILCOM 2004. Military
  Communications Conference, 2004.}, vol.~1, Oct 2004, pp. 430--435 Vol. 1.

\bibitem{a1}
A.~{Kashyap} and M.~{Shayman}, ``Routing and traffic engineering in hybrid
  rf/fso networks,'' in \emph{IEEE International Conference on Communications,
  2005. ICC 2005. 2005}, vol.~5, May 2005, pp. 3427--3433 Vol. 5.

\bibitem{back1}
G.~T. {Djordjevic}, M.~I. {Petkovic}, A.~M. {Cvetkovic}, and G.~K.
  {Karagiannidis}, ``Mixed rf/fso relaying with outdated channel state
  information,'' \emph{IEEE Journal on Selected Areas in Communications},
  vol.~33, no.~9, pp. 1935--1948, Sep. 2015.

\bibitem{back2}
Z.~{Gu}, J.~{Zhang}, and Y.~{Ji}, ``Resilience aware topology formation in
  fso-based fronthaul/backhaul networks,'' in \emph{2018 Asia Communications
  and Photonics Conference (ACP)}, Oct 2018, pp. 1--3.

\bibitem{back3}
Z.~{Gu}, J.~{Zhang}, Y.~{Ji}, L.~{Bai}, and X.~{Sun}, ``Network topology
  reconfiguration for fso-based fronthaul/backhaul in 5g+ wireless networks,''
  \emph{IEEE Access}, vol.~6, pp. 69\,426--69\,437, 2018.

\bibitem{antenna}
E.~Balti and B.~K. Johnson, ``Sub-6 ghz microstrip antenna: Design and
  radiation modeling,'' 2019.

\bibitem{saf1}
A.~J. {Torregrosa}, H.~{Maestre}, M.~L. {Rico}, E.~{Karamehmedovic}, and
  J.~{Capmany}, ``Up-conversion of eye-safe beams carrying
  2d-spatially-modulated information for detection with si-fpa cameras in fso
  applications,'' in \emph{2019 International Workshop on Fiber Optics in
  Access Networks (FOAN)}, Sep. 2019, pp. 78--82.

\bibitem{saf2}
\BIBentryALTinterwordspacing
I.~K. Son and S.~Mao, ``A survey of free space optical networks,''
  \emph{Digital Communications and Networks}, vol.~3, no.~2, pp. 67 -- 77,
  2017. [Online]. Available:
  \url{http://www.sciencedirect.com/science/article/pii/S2352864816300542}
\BIBentrySTDinterwordspacing

\bibitem{farid}
A.~A. {Farid} and S.~{Hranilovic}, ``Outage capacity optimization for
  free-space optical links with pointing errors,'' \emph{Journal of Lightwave
  Technology}, vol.~25, no.~7, pp. 1702--1710, July 2007.

\bibitem{icc}
E.~{Balti}, M.~{Guizani}, and B.~{Hamdaoui}, ``Hybrid rayleigh and
  double-weibull over impaired rf/fso system with outdated csi,'' in \emph{2017
  IEEE International Conference on Communications (ICC)}, May 2017, pp. 1--6.

\bibitem{glob16}
E.~{Balti}, M.~{Guizani}, B.~{Hamdaoui}, and Y.~{Maalej}, ``Partial relay
  selection for hybrid rf/fso systems with hardware impairments,'' in
  \emph{2016 IEEE Global Communications Conference (GLOBECOM)}, Dec 2016, pp.
  1--6.

\bibitem{glob17}
E.~{Balti}, M.~{Guizani}, B.~{Hamdaoui}, and B.~{Khalfi}, ``Mixed rf/fso
  relaying systems with hardware impairments,'' in \emph{GLOBECOM 2017 - 2017
  IEEE Global Communications Conference}, Dec 2017, pp. 1--6.

\bibitem{limit1}
N.~D. {Chatzidiamantis}, H.~G. {Sandalidis}, G.~K. {Karagiannidis}, S.~A.
  {Kotsopoulos}, and M.~{Matthaiou}, ``New results on turbulence modeling for
  free-space optical systems,'' in \emph{2010 17th International Conference on
  Telecommunications}, April 2010, pp. 487--492.

\bibitem{limit2}
M.~A. {Kashani}, M.~{Uysal}, and M.~{Kavehrad}, ``A novel statistical model for
  turbulence-induced fading in free-space optical systems,'' in \emph{2013 15th
  International Conference on Transparent Optical Networks (ICTON)}, June 2013,
  pp. 1--5.

\bibitem{mal1}
A.~{Jurado-Navas}, J.~M. {Garrido-Balsells}, J.~F. {Paris},
  M.~{Castillo-V\'azquez}, and A.~{Puerta-Notario}, ``Further insights on
  m\'alaga distribution for atmospheric optical communications,'' in \emph{2012
  International Workshop on Optical Wireless Communications (IWOW)}, Oct 2012,
  pp. 1--3.

\bibitem{mal2}
A.~Jurado-Navas, ``\BIBforeignlanguage{English}{M\'alaga statistical
  distribution: the new universal analytical propagation model for atmospheric
  optical communications},'' \emph{\BIBforeignlanguage{English}{Journal of
  Physical Chemistry \& Biophysics}}, vol.~5, no.~4, 2015.

\bibitem{beck}
\BIBentryALTinterwordspacing
R.~Boluda-Ruiz, A.~Garc\'{i}a-Zambrana, C.~Castillo-V\'{a}zquez, and
  B.~Castillo-V\'{a}zquez, ``Novel approximation of misalignment fading modeled
  by beckmann distribution on free-space optical links,'' \emph{Opt. Express},
  vol.~24, no.~20, pp. 22\,635--22\,649, Oct 2016. [Online]. Available:
  \url{http://www.opticsexpress.org/abstract.cfm?URI=oe-24-20-22635}
\BIBentrySTDinterwordspacing

\bibitem{p2}
W.~{Gappmair}, S.~{Hranilovic}, and E.~{Leitgeb}, ``Ook performance for
  terrestrial fso links in turbulent atmosphere with pointing errors modeled by
  hoyt distributions,'' \emph{IEEE Communications Letters}, vol.~15, no.~8, pp.
  875--877, August 2011.

\bibitem{p1}
F.~{Yang}, J.~{Cheng}, and T.~A. {Tsiftsis}, ``Free-space optical communication
  with nonzero boresight pointing errors,'' \emph{IEEE Transactions on
  Communications}, vol.~62, no.~2, pp. 713--725, February 2014.

\bibitem{p3}
A.~A. {Farid} and S.~{Hranilovic}, ``Diversity gain and outage probability for
  mimo free-space optical links with misalignment,'' \emph{IEEE Transactions on
  Communications}, vol.~60, no.~2, pp. 479--487, February 2012.

\bibitem{tract}
E.~{Balti} and B.~K. {Johnson}, ``Tractable approach to mmwaves cellular
  analysis with fso backhauling under feedback delay and hardware
  limitations,'' \emph{IEEE Transactions on Wireless Communications}, vol.~19,
  no.~1, pp. 410--422, Jan 2020.

\bibitem{mixed}
E.~{Balti} and M.~{Guizani}, ``Mixed rf/fso cooperative relaying systems with
  co-channel interference,'' \emph{IEEE Transactions on Communications},
  vol.~66, no.~9, pp. 4014--4027, Sep. 2018.

\bibitem{aggregate}
E.~Balti, M.~Guizani, B.~Hamdaoui, and B.~Khalfi, ``{Aggregate Hardware
  Impairments Over Mixed RF/FSO Relaying Systems With Outdated CSI},''
  \emph{IEEE Transactions on Communications}, vol.~PP, no.~99, pp. 1--1, 2017.

\bibitem{ibo}
E.~{Balti} and M.~{Guizani}, ``Impact of non-linear high-power amplifiers on
  cooperative relaying systems,'' \emph{IEEE Transactions on Communications},
  vol.~65, no.~10, pp. 4163--4175, Oct 2017.

\bibitem{v2xdiversity}
E.~Balti and B.~K. Johnson, ``Mmwaves cellular v2x for cooperative diversity
  relay fast fading channels,'' 2020.

\bibitem{asym}
------, ``Asymmetric rf/fso relaying with hpa non-linearities and feedback
  delay constraints,'' 2019.

\bibitem{eT}
E.~Balti, ``\BIBforeignlanguage{English}{Analysis of hybrid free space optics
  and radio frequency cooperative relaying systems},'' Master's thesis, 2018.

\bibitem{joint}
E.~Balti and B.~K. Johnson, ``On the joint effects of hpa nonlinearities and iq
  imbalance on mixed rf/fso cooperative systems,'' 2020.

\bibitem{variable}
L.~{Yang}, M.~O. {Hasna}, and X.~{Gao}, ``Performance of mixed rf/fso with
  variable gain over generalized atmospheric turbulence channels,'' \emph{IEEE
  Journal on Selected Areas in Communications}, vol.~33, no.~9, pp. 1913--1924,
  Sep. 2015.

\bibitem{qe}
K.~{Kumar} and D.~K. {Borah}, ``Quantize and encode relaying through fso and
  hybrid fso/rf links,'' \emph{IEEE Transactions on Vehicular Technology},
  vol.~64, no.~6, pp. 2361--2374, June 2015.

\bibitem{qf}
I.~{Avram} and M.~{Moeneclaey}, ``Quantize and forward cooperative
  communication: Joint channel and frequency offset estimation,'' in \emph{2012
  IEEE 23rd International Symposium on Personal, Indoor and Mobile Radio
  Communications - (PIMRC)}, Sep. 2012, pp. 845--850.

\bibitem{irs1}
C.~{Huang}, A.~{Zappone}, G.~C. {Alexandropoulos}, M.~{Debbah}, and C.~{Yuen},
  ``Reconfigurable intelligent surfaces for energy efficiency in wireless
  communication,'' \emph{IEEE Transactions on Wireless Communications},
  vol.~18, no.~8, pp. 4157--4170, Aug 2019.

\bibitem{irs2}
O.~{Ozdogan}, E.~{Bjornson}, and E.~G. {Larsson}, ``Intelligent reflecting
  surfaces: Physics, propagation, and pathloss modeling,'' \emph{IEEE Wireless
  Communications Letters}, pp. 1--1, 2019.

\bibitem{irs3}
E.~{Bjornson}, O.~{Ozdogan}, and E.~G. {Larsson}, ``Intelligent reflecting
  surface vs. decode-and-forward: How large surfaces are needed to beat
  relaying?'' \emph{IEEE Wireless Communications Letters}, pp. 1--1, 2019.

\bibitem{14}
S.~{Hu}, F.~{Rusek}, and O.~{Edfors}, ``Beyond massive mimo: The potential of
  data transmission with large intelligent surfaces,'' \emph{IEEE Transactions
  on Signal Processing}, vol.~66, no.~10, pp. 2746--2758, May 2018.

\bibitem{26}
X.~{Tan}, Z.~{Sun}, J.~M. {Jornet}, and D.~{Pados}, ``Increasing indoor
  spectrum sharing capacity using smart reflect-array,'' in \emph{2016 IEEE
  International Conference on Communications (ICC)}, May 2016, pp. 1--6.

\bibitem{27}
X.~{Tan}, Z.~{Sun}, D.~{Koutsonikolas}, and J.~M. {Jornet}, ``Enabling indoor
  mobile millimeter-wave networks based on smart reflect-arrays,'' in
  \emph{IEEE INFOCOM 2018 - IEEE Conference on Computer Communications}, April
  2018, pp. 270--278.

\bibitem{fd1}
E.~Balti, N.~Mensi, and S.~Yan, ``A modified zero-forcing max-power design for
  hybrid beamforming full-duplex systems,'' 2020.

\bibitem{fd2}
E.~Balti, N.~Mensi, and D.~B. Rawat, ``Adaptive gradient search beamforming for
  full-duplex mmwave mimo systems,'' 2020.

\bibitem{pls1}
N.~Mensi, D.~B. Rawat, and E.~Balti, ``Physical layer security for v2i
  communications: Reflecting surfaces vs. relaying,'' 2020.

\bibitem{pls2}
------, ``Pls for v2i communications using friendly jammer and double kappa-mu
  shadowed fading,'' 2020.

\bibitem{neji1}
N.~{Mensi}, A.~{Makhlouf}, and M.~{Guizani}, ``Incentives for safe driving in
  vanet,'' in \emph{2016 4th International Conference on Control Engineering
  Information Technology (CEIT)}, 2016, pp. 1--6.

\bibitem{neji2}
N.~{Mensi}, M.~{Guizani}, and A.~{Makhlouf}, ``Study of vehicular cloud during
  traffic congestion,'' in \emph{2016 4th International Conference on Control
  Engineering Information Technology (CEIT)}, 2016, pp. 1--6.

\bibitem{maalej}
Y.~{Maalej}, A.~{Abderrahim}, M.~{Guizani}, B.~{Hamdaoui}, and E.~{Balti},
  ``Advanced activity-aware multi-channel operations1609.4 in vanets for
  vehicular clouds,'' in \emph{2016 IEEE Global Communications Conference
  (GLOBECOM)}, 2016, pp. 1--6.

\bibitem{2}
A.~D. {Wyner}, ``Shannon-theoretic approach to a gaussian cellular
  multiple-access channel,'' \emph{IEEE Transactions on Information Theory},
  vol.~40, no.~6, pp. 1713--1727, Nov 1994.

\bibitem{9}
J.~{Xu}, J.~{Zhang}, and J.~G. {Andrews}, ``On the accuracy of the wyner model
  in cellular networks,'' \emph{IEEE Transactions on Wireless Communications},
  vol.~10, no.~9, pp. 3098--3109, Sep. 2011.

\bibitem{jeff1}
J.~G. {Andrews}, F.~{Baccelli}, and R.~K. {Ganti}, ``A tractable approach to
  coverage and rate in cellular networks,'' \emph{IEEE Transactions on
  Communications}, vol.~59, no.~11, pp. 3122--3134, November 2011.

\bibitem{jeff2}
T.~D. {Novlan}, H.~S. {Dhillon}, and J.~G. {Andrews}, ``Analytical modeling of
  uplink cellular networks,'' \emph{IEEE Transactions on Wireless
  Communications}, vol.~12, no.~6, pp. 2669--2679, June 2013.

\bibitem{22}
D.~B. {Taylor}, H.~S. {Dhillon}, T.~D. {Novlan}, and J.~G. {Andrews},
  ``Pairwise interaction processes for modeling cellular network topology,'' in
  \emph{2012 IEEE Global Communications Conference (GLOBECOM)}, Dec 2012, pp.
  4524--4529.

\bibitem{malaga}
A.~{Jurado-Navas}, J.~M. {Garrido-Balsells}, J.~F. {Paris},
  M.~{Castillo-Vázquez}, and A.~{Puerta-Notario}, ``{Further insights on
  M\'alaga distribution for atmospheric optical communications},'' in
  \emph{2012 International Workshop on Optical Wireless Communications (IWOW)},
  Oct 2012, pp. 1--3.

\bibitem{pathloss}
N.~D. Chatzidiamantis, G.~K. Karagiannidis, E.~E. Kriezis, and M.~Matthaiou,
  ``Diversity combining in hybrid {RF/FSO} systems with {PSK} modulation,'' in
  \emph{2011 IEEE International Conference on Communications (ICC)}, June 2011,
  pp. 1--6.

\bibitem{49}
G.~T. Djordjevic, M.~I. Petkovic, A.~M. Cvetkovic, and G.~K. Karagiannidis,
  ``{Mixed RF/FSO Relaying With Outdated Channel State Information},''
  \emph{IEEE Journal on Selected Areas in Communications}, vol.~33, no.~9, pp.
  1935--1948, Sept 2015.

\bibitem{beckmann}
R.~Boluda-Ruiz, A.~Garc\'{i}a-Zambrana, C.~Castillo-V\'{a}zquez, and
  B.~Castillo-V\'{a}zquez, ``{Novel approximation of misalignment fading
  modeled by Beckmann distribution on free-space optical links},'' \emph{Opt.
  Express}, vol.~24, no.~20, pp. 22\,635--22\,649, Oct 2016.

\bibitem{mmwavej}
B.~He and R.~Schober, ``{Bit-interleaved coded modulation for hybrid {RF/FSO}
  systems},'' \emph{IEEE Transactions on Communications}, vol.~57, no.~12, pp.
  3753--3763, December 2009.

\bibitem{mmwavec}
N.~D. Chatzidiamantis, G.~K. Karagiannidis, E.~E. Kriezis, and M.~Matthaiou,
  ``{Diversity Combining in Hybrid {RF/FSO} Systems with {PSK} Modulation},''
  in \emph{2011 IEEE International Conference on Communications (ICC)}, June
  2011, pp. 1--6.

\bibitem{scin}
M.~Niu, J.~Cheng, and J.~F. Holzman, ``{Error Rate Performance Comparison of
  Coherent and Subcarrier Intensity Modulated Optical Wireless
  Communications},'' \emph{J. Opt. Commun. Netw.}, vol.~5, no.~6, pp. 554--564,
  Jun 2013.

\bibitem{wolfram}
\BIBentryALTinterwordspacing
``The wolfram functions site.'' [Online]. Available:
  \url{http://functions.wolfram.com}
\BIBentrySTDinterwordspacing

\bibitem{brysh}
A.~Prudnikov and Y.~A. Brychkov, \emph{INTEGRAL AND SERIES, Volume 3, More
  Special Functions}, Computing Center of the USSR Academy of Sciences, Moscow,
  1990.

\bibitem{emil}
\BIBentryALTinterwordspacing
E.~Bj{\"{o}}rnson, {\"{O}}.~{\"{O}}zdogan, and E.~G. Larsson, ``{Intelligent
  Reflecting Surface vs. Decode-and-Forward: How Large Surfaces Are Needed to
  Beat Relaying?}'' \emph{CoRR}, vol. abs/1906.03949, 2019. [Online].
  Available: \url{http://arxiv.org/abs/1906.03949}
\BIBentrySTDinterwordspacing

\bibitem{pgfl}
S.~N. C. D. S. W. S. K.~J. Mecke, \emph{Stochastic geometry and its
  applications}, 3rd~ed., ser. Wiley series in probability and
  statistics.\hskip 1em plus 0.5em minus 0.4em\relax Wiley, 2013.

\end{thebibliography}
\end{document}